\documentclass[10pt,conference,letterpaper,compsoc]{IEEEtran}

\usepackage[english]{babel}    
\usepackage[latin1]{inputenc} 
\usepackage{float} 
\usepackage{graphicx} 
\usepackage{url} 
\usepackage{multirow} 
\usepackage{amssymb,amsmath}
\usepackage{algorithm,algorithmic}
\usepackage{array,color}
\usepackage[array,color]{colortbl}
\graphicspath{{./graphics/}} 

\newtheorem{theo}{Theorem}
\newtheorem{defi}{Definition}
\newtheorem{prop}{Proposition}
\newtheorem{exa}{Example}
\newtheorem{rem}{Remark}

{\samepage \hfill$\Box$\end{trivlist}}

\newcommand{\blinex}{\hfill\vspace{0.5mm} \hrule \hfill\vspace{0.5mm}\\}

\title{Formalization of malware through process calculi}

\author{Grégoire Jacob$^{(1/2)}$, Eric Filiol$^{(1)}$, Hervé Debar$^{(2)}$\\[0.6em]
\small $^1$ Superior School of Computing, Electronic and Automatic (ESIEA),\\ \small Operational Virology and Cryptology Lab, Laval , France\\
\small \texttt{filiol@esiea-ouest.fr}\\[0.6em]
\small $^2$ Orange Labs, Caen, France\\
\small \texttt{\{gregoire.jacob|herve.debar\}@orange-ftgroup.com}\\}

\begin{document}

\maketitle

\begin{abstract}
Since the seminal work from F. Cohen in the eighties, abstract virology has seen the apparition of successive viral models, all based on Turing-equivalent formalisms. But considering recent malware such as rootkits or k-ary codes, these viral models only partially cover these evolved threats. The problem is that Turing-equivalent models do not support interactive computations. New models have thus appeared, offering support for these evolved malware, but loosing the unified approach in the way. This article provides a basis for a unified malware model founded on process algebras and in particular the Join-Calculus. In terms of expressiveness, the new model supports the fundamental definitions based on self-replication and adds support for interactions, concurrency and non-termination allows the definition of more complex behaviors. Evolved malware such as rootkits can now be thoroughly modeled. In terms of detection and prevention, the fundamental results of undecidability and isolation still hold. However the process-based model has permitted to establish new results: identification of fragments from the Join-Calculus where malware detection becomes decidable, formal definition of the non-infection property, approximate solutions to restrict malware propagation. 
\end{abstract}
\begin{IEEEkeywords}
Malware theoretical models -- Malware detection and prevention  -- Process Algebra -- Information flow.
\end{IEEEkeywords}

\section{Introduction}

\indent Looking at recent publications, process calculi such as the $\pi$-calculus are widespread in the modeling of biological systems either molecular-based or cellular-based \cite{CCGKP08,KN06}. Computer virology is a domain where numerous parallels can be drawn between infectious diseases and malicious codes, commonly called malware. A question can be naturally raised: are process calculi also adapted to computer virology? 

\subsection{Related works and contribution}

\indent Considering malware, a recent article underlines the fact that interactions with the execution environment, concurrency and also non-termination prove to be important computation functionalities \cite{JFD07}. In effect, malware, being resilient and adaptive by nature, intensively use these functionalities to survive and infect new systems. Looking at the theoretical models existing in abstract virology, they mainly focus on the self-replication capacity which is defined in a purely functional way \cite{AD90},\cite[Chpt.2-3]{FI05},\cite{BKM06}. Unfortunately, these models rely on Turing-equivalent formalisms which can hardly support interactive computations. With the apparition of interaction-based viral techniques, new models have thus been introduced to cope with this drawback, but loosing the unified approach in the way. The apparition of k-ary malware is an obvious example. In effect, these malware heavily rely on concurrency by a distribution of the malicious code over several executing parts. A new model based on Boolean functions has been provided to model their evolving interdependence over time \cite{FI07b}. A second relevant example is the apparition of reactive non-terminating techniques such as stealth currently deployed in rootkits. Different models have been provided to cover stealth based either on steganography \cite{FIL07a} or graph theory \cite{DV08}.

\indent According to \cite{JFD07}, by evolving towards interaction-dedicated formalisms such as process calculi, a unified, reference model for malware could be defined to support these innovative techniques. Generally speaking, process calculi model the computer notion of process, that is to say an executing entity, mobile and communicating inside a context \cite{M99}. This perspective is closer to our nowadays vision of computer systems. The problem is now to choose the most adapted process calculus between the different existing ones. In order to keep the expressiveness of former models based on self-replication, the chosen process calculus must support both functional and interactive aspects. After study, the Join-Calculus was found to be the most adequate for building a malware model \cite{FO98,FG00}.

\indent As previously said, moving towards process calculi makes the malware model closer to reality while offering a greater expressiveness. However, the model still provides reasoning and proof facilities since it relies on an established theoretical formalism. But this is not the only benefit. The interactive aspects increase the visibility of computations and information flows. As a consequence, the identification of potential detection methods and the localization of possible control points become proportionally easier. 
The contribution of this article can be summed-up to the following points:
\begin{itemize}
\item Elaboration of a new viral model based on the Join-Calculus. Starting from the self-replication mechanism from functional models, this new model subsequently extends their expressiveness to support interactions, concurrency and infinite reactive computations.
\item Extension of the viral model to generic malware through a parametrization of the key components: the replication mechanism, the research of the replication target and the payload. 
\item Study of the impact of the formalism migration on the fundamental results concerning detection and prevention. 
\end{itemize}

	The article is articulated as follows. A first short introduction of the Join-Calculus is given to end this introduction. Section 2 briefly summarizes the functional notion of self-replication inside former viral models. Section 3 introduces the new process-based model which allows the definition of a distributed, context-dependent version of the self-replication. Section 4 extends the model to generic malware with an example of model parametrization to support companion viruses and rootkits. Once the model established, Section 5 addresses the existence of an algorithm either to detect malware relatively to a system context or to assess the resistance of system contexts relatively to a given class of malware. At last, Section 6 focuses on proactive solutions with the purpose of malware prevention.

\subsection{Introducing the Join-Calculus}

\indent This minimal introduction is only given to guarantee the self-containment of the article. Any reader interested in a thorough introduction is referred to the relative literature \cite{FO98,FG00}. At the basis of the Join-Calculus, an infinite set $N$ of names $x,y,z...$ is defined. Names can be compound into vectors using the notation $\overrightarrow{x}$ equivalent to $x_0,x_1,...,x_n$. Names constitute the basic blocks for message emissions of the form $x\!\!<\!\!v\!\!>$ where $x$ is called the \textit{channel} and $v$ the transmitted \textit{message}. Given in the Figure \ref{fig:syntax}, the syntax of the Join-Calculus defines three different elements to handle message passing: processes ($P$) being the communicating entities, definitions ($D$) describing the system evolution resulting of the interprocess-communication, and the join-patterns ($J$) describing the channels and messages involved in the communication \cite[pp.57-60]{FO98}.

\indent For ease of modeling, the syntactic facilities offered by the support of expressions ($E$) have been introduced \cite[pp.91-92]{FO98}. These facilities can model among others the synchronous channels necessary to concurrent functional languages. Notice that these additional facilities can be translated into the minimal core of the Join-Calculus. 
	
\begin{figure}[!h]
\begin{small}
\begin{tabular}{@{}l@{ }l@{ }l@{ }l}
$P$ & $::=$ & $v\!<\!E_1;...;E_n\!>$ & asynchronous message\\
    & $\;\mid$ & $def \; D \; in \; P$ & local definition\\
    & $\;\mid$ & $P \; \mid \; P$ & parallel composition\\ 
    & $\;\mid$ & $0$ & null process\\   
    & $\;\mid$ & $E;P$ & sequence\\  
    & $\;\mid$ & $let \; x_1,...,x_m = E \; in \; P$ & expression computation\\  
    & $\;\mid$ & $return \; E_1,...,E_n \; to \; x$ & synchronous return\\  
$E$ & $::=$ & $v(E_1;...;E_n)$ & synchronous call\\
    & $\;\mid$ & $def \; D \; in \; E$ & local definition\\
    & $\;\mid$ & $E;E$ & sequence\\  
    & $\;\mid$ & $let \; x_1,...,x_m = E \; in \; E$ & synchronous call\\ 
$D$ & $::=$ & $J \triangleright P$ & reaction rule\\
	& $\;\mid$ & $D \wedge D$ & definition conjunction\\
	& $\;\mid$ & $\top $ & null definition\\
$J$ & $::=$ & $x\!<\!y_1,...,y_n\!>$ & message pattern\\
	& $\;\mid$ & $x(y_1;...;y_n)$ & call pattern\\
	& $\;\mid$ & $J \; \mid \; J$ & join of patterns\\ 
\end{tabular}
\end{small}
\caption{Enriched syntax for the Join-Calculus. \label{fig:syntax}}
\end{figure}

	Based on the syntax, the names are divided between three sets: 1) the channels defined through a join definition ($dv$), 2) the names bound by a join-pattern ($rv$) and 3) the free names ($fv$). The inductive construction of these sets can be found in \cite[p.47]{FO98}. In addition to the syntax, an operational semantic is mandatory to establish the computational model. The semantic is established by a Reflexive Chemical Abstract Machine (RCHAM) described by the rules from the Figure \ref{fig:semantic} \cite[pp.56-62]{FO98}. In particular, the reduction rule describes the system evolution after the resolution of an exchange of messages. The reduction only occurs if the exchanged messages satisfy the join-pattern of an existing definition:\\[0.2em]
$def \; x(\overrightarrow{z}) \; \triangleright P \; in \; x(\overrightarrow{y}) \longrightarrow P\{\overrightarrow{y}/\overrightarrow{z}\}$ where $\{\overrightarrow{y}/\overrightarrow{z}\}$ is the name substitution.

\begin{figure}[!h]
\begin{small}
\begin{tabular}{lrll}
STR-JOIN  & $\vdash P_1 \; \mid \; P_2$ & $\; \rightleftharpoons $ & $\vdash P_1;P_2$\\
STR-NULL  & $\vdash 0$ & $\; \rightleftharpoons $ & $\vdash $\\
STR-AND   & $D_1 \wedge D_2 \vdash$ & $\; \rightleftharpoons$ & $D_1,D_2\vdash $\\
STR-NODEF & $T \vdash$ & $\; \rightleftharpoons $ & $\vdash $\\
STR-DEF   & $\vdash def \; D \; in \; P$ & $\; \rightleftharpoons$ & $D\sigma_{dv} \vdash P\sigma_{dv}$\\
RED       & $J\triangleright P\vdash J\sigma_{rv}$ & $\longrightarrow$ & $J\triangleright P\vdash P\sigma_{rv}$\\\\
\end{tabular}

\textbf{Substitution conditions:}\\
-STR-DEF: $\sigma_{dv}$ substitutes the defined channels from $dv[D]$ using freshly generated, distinct names.\\
-RED: $\sigma_{rv}$ substitutes the transmitted messaged to the bound names from $rv[J]$.
\end{small}
\caption{Join-Calculus Operational semantic. \label{fig:semantic}}
\end{figure}

\section{Autonomous self-replication in virology}
\label{sec:autonomous-replication}

\indent The notion of self-replication is at the heart of computer virology since it is the common denominator between the different classes of viruses and worms. Referring to the early works of J. von Neuman \cite{vN66}, two fundamental concepts are mandatory for self-replication: a replication mechanism and the existence of a self-description also called self-reference.

\indent As corroborated by successive publications \cite{KRA80,AD90},\cite[Chpt.2-3]{FI05},\cite{BKM06},  self-replication proves to be directly linked to the concept of recursion being present in the different computation paradigms. In these different functional viral models, all Turing-equivalent according to the Church-Turing thesis \cite{Rog87}, both the self-reference and the replication mechanism can be identified. Let us consider Definition \ref{defi:bkmvirus} extracted from \cite{BKM06}. This virus definition remains the most expressive and flexible viral model which actually proves to be compatible with former ones. As a consequence of Kleene's recursion theorem \cite{Rog87}, a virus is built as the solution of a fixed point equation.\\
	
\begin{defi}
\label{defi:bkmvirus}
Using a Gödel numbering, programs are indexed by integers and $\varphi _p(x)$ denotes the computation of the program indexed by $p$ over the argument $x$. According to Bonfante, Kaczmarek and Marion, a virus $v$ is a program which, for all values of $p$ and $x$ over the computation domain $D$, satisfies the equation $ \varphi _v(p,x)=\varphi _{\beta (v,p)}(x)$ where $\beta$ denotes the propagation method.\\
\end{defi}

\indent In this definition, the concepts necessary to self-replication are explicitly defined. The replication mechanism is defined through the propagation function $\beta$. As for the self-reference, it is denoted by the program $v$ which is both considered as an executed program and a parameter for the propagation function whether it is on the left or the right side of the equation. The program $p$ is called the target of the replication and the function $\beta$ implicitly contains a research routine for selecting a new valid target for the next replication. These different terms are important and must be kept in mind since they are reused all along the article.

\section{Distributed self-replication}
\indent As underlined by M. Webster in its classification \cite{WM07}, self-replicating systems, and in particular viruses, do not necessarily contain their own self-reference access or their own replication mechanism. They often rely on external services to access these fundamental elements. Let us consider an interpreted virus in bash \cite[Chpt.7]{FI05}; the replication is achieved using commands provided by the language such as \texttt{cp} and the self-reference is accessed through \texttt{\$0}. Therefore, the advantages offered by process calculi in terms of modeling become undeniable: exchanges between the process and their environment, possible distribution of the computations.\\
\indent As seen in the previous section, for self-replication to be modeled functionally, the self-reference notion is required; so it is for process modeling. In order to self-reference themselves, programs must be built as process abstractions (definition with a single pattern): $D_p = def \; p(\overrightarrow{arg})\triangleright P$ where $P$ is defined in function of the argument vector $\overrightarrow{arg}$. The program execution is therefore a process instantiation of the abstraction: $E_p = def \; D_p \; in \; p(\overrightarrow{val})$. \textbf{This hypothesis will be kept all along the article even if it is not explicitely recalled}. Based on this hypothesis, self-replication becomes the emission of this definition on an external channel, this channel being the target of the replication.\\

\begin{defi} (SELF-REPLICATION)\label{defi:self-replication} A program is said self-replicating over a channel $c$, where $c$ is the replication target, if it can be modeled as a Join-Calculus definition capable to propagate itself (i.e. to extrude itself beyond its scope) on this channel. This definition can be translated as follows: $def \; s(c,\overrightarrow{x}) \; \triangleright R$ where $R\Downarrow_{c(s)}$. $\Downarrow_{c(s)}$ is the barb predicate where the value transmitted over the channel $c$ is no longer open to any name but restricted to $v$. In this definition, $s$ denotes the self-reference whereas $R$ specifies the replication mechanism over $c$.   \\
\end{defi}

\subsection{Modeling the environment}
\label{sec:environment}
\indent Before speaking of any distribution of the replication, the execution environment in which processes evolve must be thoroughly defined. To draw a parallel with Cohen's model, a viral sequence must be considered with respect to a defined Turing Machine. If left undefined, he actually proved in \cite{CO89} that considering any sequence, a Turing Machine can be found for which this sequence is a virus.\\ 
\indent Process contexts prove to be useful tools to define execution environments. Let us consider that all execution environments share an identical global structure that can be defined as a process context. Generally speaking, a running operating system, just like any other execution machine, provides different services (system calls) and resources (memory space, files, registry). Let us define a system context denoted $C_{sys}[.]_{S\cup R}$ where services and resources constitute the common bricks, formalized by channel definitions in the Join-Calculus:\vspace*{0.4em}

\noindent \textbf{Services:} In the Join-Calculus, the available services $S$ can be modeled  by definitions with a behavior similar to execution servers waiting for queries. The services itself will be represented by a function conveyed by the variable $f_{sv}$. When the service is called, the application of $f_{sv}$ to the arguments is computed and sent back.\\[0,2em]
$\bullet$ \begin{small}$\quad def \; S_{sv}(\overrightarrow{arg}) \triangleright return \; f_{sv}(\overrightarrow{arg}) \; in \; ...$\end{small}.\vspace*{0.4em}

\noindent \textbf{Resources:} The set of resources $R$ provide storing facilities accessible to processes. Resources can be modeled by parametric processes storing information inside internal channels. Resources can be either static providing reading and writing accesses only (data file, registry keys) or executable triggered on command (programs).\\[0,2em]
$\bullet \;$ Let us consider three simple variables $c$, $c_{new}$, $c_0$:\\
\begin{small}$def \; R_{stat}(c_0) \; \triangleright$\\
$def \; (write(c_{new})|content$$<$$c$$>$$) \; \triangleright$\\
\indent \indent $(return \; to \; write |content$$<$$c_{new}$$>$$)$\\
$\wedge \; (read()|content$$<$$c$$>$$) \; \triangleright$\\
\indent \indent $(return \; c \; to \; read|content$$<$$c$$>$$)$\\
$in \; content$$<$$c_0$$>$$|return \; read,write \; to \; R_{stat} \; in \; ...$\end{small}\\[0,2em]
$\bullet \;$ Let us consider three functions $f$, $f_{new}$, $f_0$:\\
\begin{small}$def \; R_{exec}(f_0) \; \triangleright $\\
$def \; (write(f_{new})|content$$<$$f$$>$$) \; \triangleright$\\
\indent \indent $(return \; to \; write |content$$<$$f_{new}$$>$$)$\\
$\wedge \; (read()|content$$<$$f$$>$$) \; \triangleright$\\
\indent \indent  $(return \; f \; to \; read|content$$<$$f$$>$$)$\\
$\wedge \; (exec(\overrightarrow{arg})|content$$<$$f$$>$$) \; \triangleright$\\
\indent \indent $(return \; f(\overrightarrow{arg}) \; to \; exec|content$$<$$f$$>$$)$\\
$in \; content$$<$$f_0$$>$$|return \; read,write,exec \; to \; R_{exec} \; in \; ...$\\[0,2em]\end{small}

\indent This system context, split between services and resources, is compliant with the nowadays vision of computer, or more generically, with most execution environments. A process alone can not be infectious; it is viral only if the necessary services and resources to replicate are provided by the system as well as a potential external target. Considering this vision, the notion of virus can now be defined relatively to a system context by construction of the viral sets \cite{CO89}.

\subsection{Construction of the viral sets}
\indent Program replication is formalized by the emission of its definition on an external channel provided by the environment. Consequently, the barb predicate defined in the different process calculi is unadapted: transmitted values are omitted and once the program is placed inside a process context, reactions become internal and thus no longer observable by a barb predicate. We will thus define a new predicate, more adapted, that will be called valued-reaction. Its definition is given below. \\

\begin{defi} (VALUED-REACTION) Let $P$, $P'$ be two processes, $x$ a channel and $a$ a value from $P$ (either bound or free). A valued-reaction $P\nabla_{x(a)}P'$ occurs if and only if $P = C[x\!\!<\!\!a\!\!>]_S$ for some process context $C[.]_S$ capturing $x$ i.e. $x \!\in\! S$. By reduction on join pattern $x\!\!<\!\!a\!\!>$, $P\longrightarrow P'$. The $\nabla_{x(a)}$ predicate syntactically checks the possibility for a process $P$ to react on a channel $x$ with the value $a$. Once resolved, the reaction leads to a new process state $P'$.\\
\end{defi}

\indent Using valued-reaction, we can now define the principle of viable replication in a given environment. Viable replication guarantees that the replicated version of a program is still capable of self-replication. This principle was already present in the self-reproducing cellular automata from J. von Neuman where cellular configurations iteratively rebuild themselves at each transition \cite{vN66}. Similarly, the replication is iterated by valued-reactions through two phenomenons:\\
\textbf{-Original replication:} During the first execution of the program $p$, denoted by the process $P$, $p$ is replicated over a resource channel. This channel is consumed by the system context to evolve towards a new state. This is represented by the predicate:\\[0.2em]
\begin{small}$\exists x\in R,C[P]_{S\cup R}\nabla_{x(p)}P'$\end{small}.\\[0.2em]
\textbf{-Successive replications:} The successive iterations of the replications are triggered by the activation of the intermediate infected resources. If $P^{(i)}$ corresponds to the execution of the $i^{th}$ infected form, then, the following predicate should be verified:\\[0.2em]
\begin{small}$\exists x\in R^{(i)},C^{(i)}[P^{(i)}]_{S\cup R^{(i)}}\nabla_{x(p)}P^{(i+1)}$\end{small}.\\[0.2em]
\indent By definition, the viral sets contain the processes satisfying the viable self-replication principle. The notion of viral set from F. Cohen must thus be extended relatively to a system context which conditions the consumption of the replicated definitions and the activation of the intermediate infected forms. In fact, these viral sets can be built following the same method of iterated replications.  \\



\begin{defi}(VIRAL SET)  Let us consider a system context $C_{sys}[.]_{S\cup R}$ where $S$ denotes the available services and $R$ the accesses to resources. Its viral set $E_v$ can be recursively constructed as follows. \\[0,2em]
\begin{small}
$E_v(C_{sys}[.]_{S\cup R}) = \{V \; | \; \exists \overrightarrow{x} \; of \; size \; n\geq 2$\\[0,2em]
\indent $\exists \overrightarrow{x} \; of \; size \; n\geq 2 \; and \; \overrightarrow{exec} \; of \; size \; n\!-\!1 \; such \; as$\\[0,2em]
\indent $C_{sys}[V]_{S\cup R}\nabla_{v(y_0)}\nabla_{x_0(v)}C'_{sys}[V']_{S\cup R'}$\\[0,2em]
\indent and for all $1 \leq i < n$,\\[0,2em]
\indent $C^{(i)}_{sys}[exec_i\!<\!\overrightarrow{arg_i}\!>]_{S\cup R^{(i)}}\nabla_{v(y_i)}\nabla_{x_i(v)}$\\[0,2em]
\indent $C^{(i+1)}_{sys}[V^{(i+1)}]_{S\cup R^{(i+1)}}$\\[0,2em]
$\}$\\[0,2em]
\end{small}
with the following constraints:\\
- All $x_i \!\in\! \overrightarrow{x}$ denote replication targets. They can be either channels to existing resources: $x_i \!\in\! R$, or to dynamically created resources: $x_i \!\in\! R^{(i)}$ meaning that $x_i \!\in\! rv(J)$ where $J$ is a join related to a resource definition $D$ with $dv(D) \subset R$,\\
- The messages $exec_i\!<\!\overrightarrow{arg_i}\!>$ activate the intermediate infected resources. To avoid biases, they must not simulate viral activity : $exec_i \!\in\! R^{(i)}$ and $arg_i \!\not \in\! bv(V)$.
\end{defi}

\subsection{Distributed virus replication}
\label{subsec:parametric-virus-replication}

\subsubsection{Environment refinement for replication}
\indent Considering self-replication, several services and resources must be defined because they may be externalized by the self-reproducing systems \cite{WM07}: access to the self-reference, replication mechanisms and replication targets. The structure for services and resources, globally defined in the system context from Section \ref{sec:environment}, must thus be refined to support these features. The refined definitions are given below with relevant examples from current operating systems given in the Table \ref{tab:osservices}: 
\begin{description}[\IEEEsetlabelwidth{$\;\;$}]
\item[\textbf{Self-reference access:}] $\qquad \qquad \qquad \qquad$ Today's operating systems all handle a list of executing processes, with a particular pointer on the active process. This list is among others used for scheduling. A service is often provided to access this list and in particular the pointed active process which denotes the self-reference. In order to maintain this list, the program executions must be launched through a dedicated service.\\
\begin{small}
$\bullet \; D_{proc} \stackrel{\scriptscriptstyle \rm def}{=} proc_{exec}(p,\overrightarrow{args}) \; \triangleright$\\
$sys_{updt}(p). return \; p(\overrightarrow{args}) \; to \; proc_{exec}$\\
$\bullet \; D_{ref} \stackrel{\scriptscriptstyle \rm def}{=} (sys_{updt}(r_{new})|current$$<$$r_{cur}$$>$$) \; \triangleright $\\
\indent \indent $current\!\!<\!\!r_{new}\!\!>$\\
$\wedge \; (sys_{ref}()|current$$<$$r_{cur}$$>$$) \; \triangleright$\\
\indent \indent $(current$$<$$r_{cur}$$>$$|return \; r_{cur} \; to \; sys_{ref})$\end{small}\\
Self-reference access must be considered as a service even if it uses an internal resource. A solution is to publish only $sys_{ref}$ and $proc_{exec}$ in $S$ (from $C_{sys}[.]_{S\cup R}$). Any process placed in the context will not have direct writing access to the internal channel $current$ storing the reference. From the process perspective, the two provided channels will be similar to services .
\item[\textbf{Replication mechanism:}] $\qquad \qquad \qquad \qquad \quad$ The replication mechanism is a function $r$ which copies data from an input channel towards and output channel. The function $r$ has been deliberately left parametric for the model to remain generic. However $r$ is strongly constrained to forward the input data towards the output channel after an indefinite number of transformations.\\
\begin{small}
$\bullet \; D_{rep} \stackrel{\scriptscriptstyle \rm def}{=} sys_{rep}(in,out)\; \triangleright$\\
$return \; r(in,out) \; to \; sys_{rep}$\end{small}.
\item[\textbf{Replication targets:}] $\qquad \qquad \qquad \quad \;$ A pool of executable resources constitute the replications targets. These resources can be preexisting (infection) or created by the malware (duplication).\\
\begin{small}
$\bullet \; D_{targ} \stackrel{\scriptscriptstyle \rm def}{=} R_{targ}(f_{init}) \; \triangleright $\\
$def \; (write(f_{new})|content$$<$$f$$>$$) \; \triangleright$\\
$(return \; to \; write|content$$<$$f_{new}$$>$$)$\\
$\wedge \; (read()|content$$<$$f$$>$$) \; \triangleright$\\
$(return \; f \; to \; read|content$$<$$f$$>$$)$\\
$\wedge \; (exec(\overrightarrow{arg})|content$$<$$f$$>$$) \; \triangleright$\\
$(return \; proc_{exec}(f,\overrightarrow{arg}) \; to \; exec|content$$<$$f$$>$$) \; in$\\
$content$$<$$f_{init}$$>$$|return \; read,write,exec\; to \;R_{targ}$\end{small}
\end{description}\vspace*{0.3em}
Using the previous definition, a system with $n$ resources can be defined as an evaluation context. This context being enough generic to applied to the majority of existing systems, we will consider this system context all along this section for the different definitions and proofs:\\
$C_{sys}[.]_{S\cup R} \stackrel{\scriptscriptstyle \rm def}{=} def \; D_{proc} \wedge  D_{ref} \wedge  D_{rep} \wedge D_{targ} \; in$\\
$let \; sr_1,sw_1,se_1,...,sr_n,sw_n,se_n = $\\
$R_{targ}(f_{1}),...,R_{targ}(f_{n}) \; in \; (current$$<$$null$$>$$ \, | \, [.])$\\
where:\\
$S=\{proc_{exec},sys_{ref},sys_{rep}\}$ and $R=\{\overrightarrow{sr},\overrightarrow{sw},\overrightarrow{se}\}$

\begin{table}[!ht]
\centering
\begin{footnotesize}
\begin{tabular}{|l|l|l|}
\hline
\multicolumn{3}{|l|}{Services provided by well-known operating systems}\\
\hline
\rowcolor[gray]{0} \textcolor{white}{Channels} & \textcolor{white}{Linux APIs} & \textcolor{white}{Windows APIs}\\
\hline
$proc_{exec}$ & fork( ), exec( ) & CreateProcess( )\\
\hline
$sys_{ref}$ & getpid( ), & GetCurrentProcess( ), \\
  & readlink( ) & GetModuleFileName( ) \\
\hline
$sys_{rep}$ & sendfile( ) & CopyFile( )  \\
\hline
$Resources$ & fread( ), & ReadFile( ), \\
$Accesses$ & fwrite( ), &  WriteFile( ),  \\
  & ... & ...  \\
\hline
\end{tabular}
\end{footnotesize}
\caption{Parallel between refined channels and equivalent OS services and resource accesses.\label{tab:osservices}}
\end{table}

\subsubsection{Classes of self-replicating  viruses}
\indent Using this refined system context, the four classes of self-replicating viruses from M. Webster \cite{WM07} can be defined in this process-based model. Through these four classes, the important components required for autonomous replication can be found (see Section \ref{sec:autonomous-replication}): the access to the self-reference, a replication mechanism denoted by the function $r$ and a target research routine denoted by the function $t$. These two last functions have been willingly left parameterizable. 

\indent Through parametrization, several types of replication can be supported, for example: 
(1) \textit{overwriting infections} which can be defined by \begin{small}$def \; r(v,sw) \triangleright sw(v)$\end{small}, 
(2) \textit{append infections} (respectively \textit{prepend infections}) with a definition of the form \begin{small}$def \; r(v,sw,sr) \; \triangleright \; (let \; p = sr() \; in \; def \; p_1(\overrightarrow{arg}) \, \triangleright \, v().p(\overrightarrow{arg}) \; in \; sw(p_1))$\end{small}, 
(3) \textit{companion infections} described in a coming section because of their greater modeling complexity.

\indent With regards to the concept of self-replication from Definition \ref{defi:self-replication}, the virus case is particular since the target of the replication is no longer passed as a parameter but chosen by an internal research routine. The behavior of this routine, just like the replication mechanism, must remain parameterizable. Generally speaking, successive replications follow three main schemes: 
(1) targets are \textit{hard-coded} in the virus, like a predefined file path for example, meaning that the target channel will always be the same $n$ such as \begin{small}$def \; t() \; \triangleright \; return \; n \; to \; t$\end{small}, 
(2) target resources are \textit{dynamically created} by the routine using the facilities of the system \begin{small}$def \; t() \; \triangleright \; let \; sr,sw,se = R(empty) \; in \; return \; sw \; to \; t$\end{small}, 
(3) target are \textit{discovered} by running through the system searching for vulnerable resources. Directory exploration is a typical example. Once again this example is too complex to be briefly described here. 
The target research must be integrated in the virus definition, in addition to the self-reference access and the replication mechanism. Based on this parametric approach, as well as on the provided modeling of the system context, four main classes of viruses can be defined according to the exported services.\\

\begin{defi}\label{def:virus-classes} Let $V$ be a viral process. Let $R$ and $S$ be the definition of sub-processes responsible for the self-reference access and and the replication mechanism. An additional definition $T$ is responsible for researching the target of the infection. At last, a process $P$ is introduced for the post-infection process i.e. the payload: 
\begin{itemize}
\begin{small}
\item $R \stackrel{\scriptscriptstyle \rm def}{=} loc_{rep}(in,out)\; \triangleright return \; r(in,out) \; to \; loc_{rep}$\end{small} where $r$ is a parametric function defining the replication mechanism.
\item \begin{small}$S \stackrel{\scriptscriptstyle \rm def}{=} loc_{ref}()\; \triangleright \; return \; v \; to \; loc_{ref}$\end{small}.
\item \begin{small}$T \stackrel{\scriptscriptstyle \rm def}{=} loc_{targ}()\; \triangleright return \; t() \; to \; loc_{rep}$\end{small} where $t$ is a parametric function defining the routine for target research.
\item $P$ is any process modeling a payload. 
\end{itemize}
Four classes of worms can be defined using these primitives and the system services:
\begin{itemize}
\item \textbf{(Class I)} V is totally autonomous:\\
\begin{small}
$V_{I} \stackrel{\scriptscriptstyle \rm def}{=} def_v \; v(\overrightarrow{x}) \triangleright (def_v \; S \wedge R \wedge T$\\
$in \; loc_{rep}(loc_{ref}(),loc_{targ}()).P) \; in \; proc_{exec}(v,\overrightarrow{a})$
\end{small}
\item \textbf{(Class II)} V uses an external replication mechanism provided by the system:\\
\begin{small}
$V_{II} \stackrel{\scriptscriptstyle \rm def}{=} def_v \; v(\overrightarrow{x}) \triangleright (def_v \; S \wedge T$\\
$in \; sys_{rep}(loc_{ref}(),loc_{targ}()).P) \; in \; proc_{exec}(v,\overrightarrow{a})$
\end{small}
\item \textbf{(Class III)} V uses an external access to the self-reference provided by the system:\\
\begin{small}
$V_{III} \stackrel{\scriptscriptstyle \rm def}{=} def_v \; v(\overrightarrow{x}) \triangleright (def \; R \wedge T$\\
$in \; loc_{rep}(sys_{ref}(),loc_{targ}()).P)  \; in \; proc_{exec}(v,\overrightarrow{a})$
\end{small}
\item \textbf{(Class IV)} V uses only external services:\\
\begin{small}
$V_{IV} \stackrel{\scriptscriptstyle \rm def}{=} def_v \; v(\overrightarrow{x}) \triangleright (def \; T$\\
$in \; sys_{rep}(sys_{ref}(),loc_{targ}()).P) \; in \; proc_{exec}(v,\overrightarrow{a})$\\
\end{small}
\end{itemize}
\end{defi}

\indent In this definition, the research routine $T$ is always internal to the virus. However, the definition would support the distribution of this functionality. This case has not been included in the definition because, to our knowledge, no malware completely externalize this functionality. On the other hand, since it runs through the environment, the target research is likely to use intensively the system services.\\ 

\begin{prop}
If the system context $C_{sys}[.]_{S\cup R}$ provides the right services and valid targets, the four virus classes $V_{I}$,$V_{II}$,$V_{III}$ and $V_{IV}$ achieve viable replication i.e. these classes are included in the viral set $E_v(C_{sys}[.]_{S\cup R})$.\\
\end{prop}

\begin{IEEEproof}
Let us consider a system context with several potential resources as defined in this section. Let us consider a simple case of parameterization for the replication mechanism $r$ and the target research $t$. Notice that other definitions could be used without modifying the core of the proof: additional reductions would only be necessary.\\[0.3em]
\begin{small}
$def \; r(x,w) \triangleright w(x)$\\
$def \; t() \triangleright return \; sw_i \; to \; t \; at \; the \; i^{th} \; iteration$.\\[0.3em]
\end{small}
\indent Let us consider the case of third class of virus with the following notations:\\[0.3em] 
\begin{small}
$D_{V_{III}} \stackrel{\scriptscriptstyle \rm def}{=} v() \; \triangleright$\\
\indent \indent \indent $(def \; R \wedge T \; in \; loc_{rep}(sys_{ref}(),loc_{targ}());P)$.\\
$D_{R_k} \stackrel{\scriptscriptstyle \rm def}{=} (sw_k(f_{new})|content_k$$<$$f$$>$$) \; \triangleright$\\
\indent \indent \indent $(content_k$$<$$f_{new}$$>$$)$\\
$\wedge \; (sr_k()|content_k$$<$$f$$>$$) \; \triangleright$\\
\indent \indent \indent $(content_k$$<$$f$$>$$|return \; f \; to \; sr_k)$\\
$\wedge \; (se_k(\overrightarrow{arg})|content_k$$<$$f$$>$$) \; \triangleright$\\
\indent \indent \indent $(content_k$$<$$f$$>$$|return \; proc_{exec}(f,\overrightarrow{arg}) \; to \; se_k)$.\\
\end{small}

\noindent To prove viable replication, it must be proven that the viral function $v$ initially infect a resource, but also that an execution request $se_1(\overrightarrow{a_1})$ reproduces the infection towards a second writing channel $sw_2$. Next iterations can then be reduced to these two cases:\\ 
\textbf{Initial infection:}\\
\begin{small}$C_{sys}[V_{III}]_{S\cup R}\nabla_{v()}\nabla_{sw_1(p)}C'_{sys}[P]_{S\cup R}$\end{small}. \\
\textbf{Successive infections:}\\
\begin{small}$C'_{sys}[se_1(\overrightarrow{a_1})]_{S\cup R}\nabla_{v()}\nabla_{sw_2(p)}C''_{sys}[P]_{S\cup R}$\end{small}. \\

\begin{footnotesize}
\noindent $\vdash C_{sys}[V_{III}]_{S\cup R}$\\
\noindent $\rightleftharpoons$ (str-def+str-and)\\[-2mm]
\blinex
\noindent $D_{proc},D_{ref},  D_{rep}, D_{targ} \vdash \\
let \; sr_1,sw_1,se_1,...,sr_n,sw_n,se_n= R_{targ}(f_{1}),...,R_{targ}(f_{n})$\\
$in \; (current$$<$$null$$>$$ \; | \; V_{III})$\\
\noindent $\longrightarrow $ (react+str-def+str-and)\\[-2mm]
\blinex
\noindent $D_{proc},D_{ref}, D_{rep}, D_{targ}, D_{R_1},...,D_{R_n} \vdash \\
content_1$$<$$f_{1}$$>$$\; | \; \Pi_{i=2}^{n} content_i$$<$$f_{i}$$>$$ \; | $\\
$current$$<$$null$$>$$ \; | def_v \; D_{V_{III}} \; in \; proc_{exec}(v,\overrightarrow{a})$\\
\noindent $\longrightarrow $ (str-def)\\[-2mm]
\blinex
\noindent $D_{proc}, D_{ref},  D_{rep}, D_{targ}, D_{R_1},...,D_{R_n}, D_{V_{III}} \vdash_{\{v\}} \\
content_1$$<$$f_{1}$$>$$\; | \; \Pi_{i=2}^{n} content_i$$<$$f_{i}$$>$$ \; |$\\
$current$$<$$null$$>$$ \; | \; proc_{exec}(v,\overrightarrow{a})$\\
\noindent $\longrightarrow $ (react)\\[-2mm]
\blinex
\noindent $D_{proc}, D_{ref},  D_{rep}, D_{targ}, D_{R_1},...,D_{R_n}, D_{V_{III}} \vdash_{\{v\}} \\
content_1$$<$$f_{1}$$>$$\; | \; \Pi_{i=2}^{n} content_i$$<$$f_{i}$$>$$ \; |$\\
$current$$<$$null$$>$$ \; | \; sys_{updt}(v).v(\overrightarrow{a})$\\
\noindent $\longrightarrow $ (react)\\[-2mm]
\blinex
\noindent $D_{proc}, D_{ref},  D_{rep}, D_{targ}, D_{R_1},...,D_{R_n}, D_{V_{III}} \vdash_{\{v\}} \\
content_1$$<$$f_{1}$$>$$\; | \; \Pi_{i=2}^{n} content_i$$<$$f_{i}$$>$$ \; | \; current$$<$$v$$>$$ \; | \; v(\overrightarrow{a})$\\
\noindent $\longrightarrow $ (react)\\[-2mm]
\blinex
\noindent $D_{proc},D_{ref},  D_{rep}, D_{targ}, D_{R_1},...,D_{R_n}, D_{V_{III}} \vdash_{\{v\}} \\
 content_1$$<$$f_{1}$$>$$\; | \; \Pi_{i=2}^{n} content_i$$<$$f_{i}$$>$$ \; | \; current$$<$$v$$>$$ \; | $\\
$def R \wedge T \; in \; loc_{rep}(sys_{ref}(),loc_{targ}()).P$\\
\noindent $\rightleftharpoons$ (str-def+str-and)\\[-2mm]
\blinex
\noindent $D_{proc},D_{ref},  D_{rep}, D_{targ}, D_{R_1},...,D_{R_n}, D_{V_{III}}, R, T \vdash_{\{v\}}\\
 content_1$$<$$f_{1}$$>$$\; | \; \Pi_{i=2}^{n} content_i$$<$$f_{i}$$>$$ \; | \; current$$<$$v$$>$$ \; | $\\
$loc_{rep}(sys_{ref}(),loc_{targ}()).P$\\
\noindent $\longrightarrow $ (react)\\[-2mm]
\blinex
\noindent $D_{proc},D_{ref},  D_{rep}, D_{targ}, D_{R_1},...,D_{R_n}, D_{V_{III}}, R, T \vdash_{\{v\}} \\
 content_1$$<$$f_{1}$$>$$\; | \; \Pi_{i=2}^{n} content_i$$<$$f_{i}$$>$$ \; | \; current$$<$$v$$>$$ \; | $\\
$loc_{rep}(v,loc_{targ}()).P$\\
\noindent $\longrightarrow $ (react)\\[-2mm]
\blinex
\noindent $D_{proc},D_{ref},  D_{rep}, D_{targ}, D_{R_1},...,D_{R_n}, D_{V_{III}}, R, T \vdash_{\{v\}} \\
 content_1$$<$$f_{1}$$>$$\; | \; \Pi_{i=2}^{n} content_i$$<$$f_{i}$$>$$ \; | \; current$$<$$v$$>$$ \; | $\\
$loc_{rep}(v,sw_1).P$\\
\noindent $\longrightarrow $ (react)\\[-2mm]
\blinex
\noindent $D_{proc},D_{ref},  D_{rep}, D_{targ}, D_{R_1},...,D_{R_n}, D_{V_{III}},R,T \vdash_{\{v\}} \\
 content_1$$<$$f_{1}$$>$$\; | \; \Pi_{i=2}^{n} content_i$$<$$f_{i}$$>$$ \; | \; current$$<$$v$$>$$ \; | $\\
$sw_1(v).P$\\
 \noindent $\longrightarrow $ (react)\\[-2mm]
\blinex
\noindent $D_{proc},D_{ref},  D_{rep}, D_{targ}, D_{R_1},...,D_{R_n}, D_{V_{III}}, R, T \vdash_{\{v\}} \\
 content_1$$<$$v$$>$$\; | \; \Pi_{i=2}^{n} content_i$$<$$f_{i}$$>$$ \; | \; current$$<$$v$$>$$ \; | \; P$\\
\end{footnotesize}

Once the initial replication is achieved, the second replication is activated from the current state thanks to an execution request $se_1(\overrightarrow{a_1})$.\\

\begin{footnotesize}
\noindent $D_{proc},D_{ref},  D_{rep}, D_{targ}, D_{R_1},...,D_{R_n}, D_{V_{III}}, R, T \vdash_{\{v\}} \\
 content_1$$<$$v$$>$$\; | \; content_2$$<$$f_{2}$$>$$ \; | \; \Pi_{i=3}^{n} content_i$$<$$f_{i}$$>$$ \; | $\\
$current$$<$$v$$>$$ \; | \; se_1(\overrightarrow{a_1})$\\
\noindent $\longrightarrow $ (react)\\[-2mm]
\blinex
\noindent $D_{proc},D_{ref},  D_{rep}, D_{targ}, D_{R_1},...,D_{R_n}, D_{V_{III}}, R, T  \vdash_{\{v\}} \\
 content_1$$<$$v$$>$$\; | \; content_2$$<$$f_{2}$$>$$ \; | \; \Pi_{i=3}^{n} content_i$$<$$f_{i}$$>$$ \; | $\\
$current$$<$$v$$>$$\; | \; proc_{exec}(v,\overrightarrow{a_1})$\\
\end{footnotesize}

From there the reduction is identical to the previous one except for the call to $loc_{targ}$ which is reduced to $sw_2$ and no longer $sw_1$.\\

\begin{footnotesize}
\noindent ...\\\blinex
\noindent $D_{ref},  D_{rep}, D_{targ}, D_{R_1},...,D_{R_n}, D_{V_{III}}, R, T, R', T'  \vdash_{\{v\}} \\
 content_1$$<$$v$$>$$\; | \; content_2$$<$$v$$>$$ \; | \; \Pi_{i=3}^{n} content_i$$<$$f_{i}$$>$$ \; |$\\
$current$$<$$v$$>$$\; | \; P \;$\\
\end{footnotesize}

These two reduction prove the viable replication for viruses of the class $V_{III}$. An identical approach can be used to provide proofs for the remaining classes. 
\end{IEEEproof}


\subsection{Distributed worm propagation}

	The propagation mechanism for worms is similar to virus replication. The difference lies in the scope of the extrusion: the abstract definition of the worm is no longer extruded to a local resource through a writing channel, but to a remote system context. This topology can be defined as contexts imbricated on two levels. A first local system context, similar to the one from the Section \ref{subsec:parametric-virus-replication}, is included into a global architectural context containing parallel remote systems and communications facilities between them (a computer network topology for example):\\[0.3em]
\noindent \textbf{Local context:} Let us define a new propagation service in the local context. The principle of the propagation service is similar to replication (the propagation function $p$ replaces the funtion $r$). This new local context can be simplified by removing the resource definitions used to store the replicated code:\\[0,3em] 
\begin{small}
$D_{prop}\stackrel{\scriptscriptstyle \rm def}{=} sys_{prop}(in,out)\; \triangleright return \; p(in,out) \; to \; sys_{prop}$\\
$C_{lsys}\stackrel{\scriptscriptstyle \rm def}{=} def \; D_{proc} \wedge  D_{ref} \wedge  D_{prop}$\\
$in \; (current$$<$$null$$>$$ \; | \; [.])$\\[0,3em] 
\end{small}
\noindent \textbf{Remote context:} The remote context must provide communication facilities between the different systems. The $ComChannel$ definition enables the generation of two-way communication channels. Processing of the data transmitted by the local context is delegated  to the remote parallel contexts running inside the global architecture. In order to simplify the model, the definition below only considers a single process $P_{rsys}$ modeling the remote system but several systems can run in parallel. In addition, the resources and services from $P_{rsys}$ can also be refined:\\[0,3em] 
\begin{small}
$P_{rsys} \stackrel{\scriptscriptstyle \rm def}{=} let\; d\!=\!rcv() \; in \;P_{processing}$\\
$C_{garch} \stackrel{\scriptscriptstyle \rm def}{=} def \; ComChannel() \; \triangleright$ \\
\indent $def \; send$$<$$m$$>$$|receive() \triangleright return \; m \; to \; receive \; in$\\
\indent $return \; send,receive \; in \; let \; sd,rcv \! = \! ComChannel()$\\
$in \; [\,P_{rsys} \; | \; [.] \;]$\\\\
\end{small}

\begin{defi} Let $W$ be a worm able to propagate to remote system using $P$, $S$ et $T$, the definitions of three sub-processes respectively responsible for propagation (pending of the replication for viruses), access to the self-reference and the research of a potential target:
\begin{itemize}
\begin{small}
\item $P \stackrel{\scriptscriptstyle \rm def}{=} loc_{prop}(in,out)\; \triangleright \; return \; p(in,out) \; to \; loc_{prop}$
\item $S \stackrel{\scriptscriptstyle \rm def}{=} loc_{ref}()\; \triangleright \; return \; w \; to \; loc_{ref}$
\item $T \stackrel{\scriptscriptstyle \rm def}{=} loc_{targ}()\; \triangleright \; return \; t() \; to \; loc_{targ}$
\end{small}
\end{itemize}
Four classes of worms can be defined using these primitives and the system services:
\begin{itemize}
\item \textbf{(Class I)} W is totally autonomous:\\
\begin{small}
$W_{I} \stackrel{\scriptscriptstyle \rm def}{=} def \; w(\overrightarrow{x}) \triangleright (def \; S \wedge P \wedge T \; in$\\
$loc_{prop}(loc_{ref}(),loc_{targ}()).P') \; in \; proc_{exec}(w,\overrightarrow{a})$
\end{small}
\item \textbf{(Class II)} W uses an external propagation mechanism provided by the system:\\
\begin{small}
$W_{II} \stackrel{\scriptscriptstyle \rm def}{=} def \; w(\overrightarrow{x}) \triangleright (def \; S \wedge T \; in$\\
$sys_{prop}(loc_{ref}(),loc_{targ}()).P') \; in \; proc_{exec}(w,\overrightarrow{a})$
\end{small}
\item \textbf{(Class III)} W uses an external access to the self-reference provided by the system:\\
\begin{small}
$W_{III} \stackrel{\scriptscriptstyle \rm def}{=} def \; w(\overrightarrow{x}) \triangleright (def \; P \wedge T \; in$\\
$loc_{prop}(sys_{ref}(),loc_{targ}()).P')  \; in \; proc_{exec}(w,\overrightarrow{a})$
\end{small}
\item \textbf{(Class IV)} W uses only external services:\\
\begin{small}
$W_{IV} \stackrel{\scriptscriptstyle \rm def}{=} def \; w(\overrightarrow{x}) \triangleright (def \; T \; in$\\
$sys_{prop}(sys_{ref}(),loc_{targ}()).P') \; in \; proc_{exec}(w,\overrightarrow{a})$\\
\end{small}
\end{itemize}
\end{defi}

	The four classes of worms satisfy viable replication just like viruses do. The main difference comes from the extrusion of the $w$ definition which is no longer bound to the local system but can be extended to the remote context.\\

\begin{rem}
Just like replication, the propagation function can be refined for more complexity. The simplest case remains the simple copy:\\
\begin{small}
$D_{prop} \stackrel{\scriptscriptstyle \rm def}{=} def \; p(in,out) \triangleright out\!<\!\!in\!\!>$\\
\end{small}
For more complex cases such as Email-worms, intermediate functions can be introduced with their counterparts in the remote system to reverse the processing:\\
\begin{small}
$D_{prop} \stackrel{\scriptscriptstyle \rm def}{=} def \; p(in,out) \; \triangleright $\\
$out\!<\!\!concat(SMTPheader,base64(in))\!\!>$\\
$P_{rsys} \stackrel{\scriptscriptstyle \rm def}{=} let \; d = rcv() \; in \; base64decode(body(d))$\\
\end{small}
The research routine $t()$ can be defined accordingly to parse the address books of different mail clients.
\end{rem}

\section{Modeling complex malicious behaviors}
	Modeling complex behaviors proves the interest of the parametric approach. This section gives examples of complex refinements both for the replication function $r$ and the payload process $P$ from the previous section.
	
\subsection{Companion viruses}
	Companion viruses remain a particular case of the parametric definition of the Section \ref{subsec:parametric-virus-replication}. Their specificity lies in their replication mechanism: instead of overwriting or modifying the content of the resource targeted by the infection, the virus replaces this resource from the system perspective. Companion viruses can be divided between two classes whether the replacement is achieved (a) by diverting the file system naming mechanism or (b) by diverting the hierarchy of execution \cite[Chpt.8]{FI05}. The replication function is consequently more complex and requires three steps:\\
\noindent 1-a) Renaming or relocation the target of the infection.\\
\noindent 1-b) Modification of the system hierarchy of execution.\\
\noindent 2) Creation of a new resource under the target name.\\
\noindent 3) Copy of the viral code in the replacing resource.\\ 

\noindent\textbf{Modeling the file system}\\
\indent In order to model a companion virus, it becomes necessary to introduce a refined model for the file system. The purpose of the file system is to associate a resource name (a system path) with a given location and access channels (reading, writing, execution). The principle is thus compatible with our model of executable resources. In addition, a file system is introduced into the environment defined in \ref{subsec:parametric-virus-replication} which is responsible for maintaining a list of $4$-tuples associated to the different files. Let us give a first definition of a file entry as well as its access and update methods:\\[0,3em] 
\begin{small}
$E_{FS} \stackrel{\scriptscriptstyle \rm def}{=} E(n_{init},sr_{init},sw_{init},se_{init}) \; \triangleright $\\
\indent $def \; n_{init}(c,p) \; | entry$$<$$sr,sw,se$$>$$ \; \triangleright $\\
\indent  \indent $(if \; [c = dl] \; then \; 0 \; else$\\
\indent  \indent $\: if \; [c = mv] \; then \; E(p,sr,sw,se) \; else$\\
\indent  \indent $\: if \; [c = ex] \; then \; se(p) \; | \; entry$$<$$sr,sw,se$$>$$ \; else$\\
\indent  \indent $\: if \; [c = rd] \; then \; p(sr()) \; | \; entry$$<$$sr,sw,se$$>$$ \; else$\\
\indent  \indent $\: if \; [c = wr] \; then \; sw(p) \; | \; entry$$<$$sr,sw,se$$>$$)$\\
\indent $in \; entry$$<$$sr_{init},sw_{init},se_{init}$ $\!\!>$ \\[0,3em]
\end{small}
\indent The file system provides different commands to manage these entries. The different command takes the file name in input, and the file system is responsible for executing these commands on the right resource (which are basically modeled as executing processes):\\
\noindent -$new$ to create new files,\\
\noindent -$delete$ to delete existing files,\\
\noindent -$move$ to modify the name of the file (modifying the name only corresponds to a renaming operation whereas modifying the complete path is a relocation),\\
\noindent -$execute$ to execute a given file,\\
\noindent -$read$ to read from a given file,\\
\noindent -$write$ to write to a given file.\\
\noindent These commands of the file system are modeled as definitions whereas the entries of the file system constitute a set of parallel processes. A file system definition is given below where the executing parallel processes correspond to the already existing files referred by the name vector $\overrightarrow{n}$:\\[0,3em] 
\begin{small}
$M_{FS} \stackrel{\scriptscriptstyle \rm def}{=} def \; E_{FS} \; in$ \\
$def \; new(n_{new}) \; \triangleright \; E(n_{new},R_{exec}(null))$\\
\indent $\quad \wedge \; delete(n_{del}) \; \triangleright \; n_{del}(dl,null)$ \\
\indent $\quad \wedge \; move(n_{old},n_{new}) \; \triangleright \; n_{old}(mv,n_{new})$ \\
\indent $\quad \wedge \; execute(n_{exe},arg) \; \triangleright \; n_{exe}(ex,arg)$ \\
\indent $\quad \wedge \; read(n_{rd,buffer},arg) \; \triangleright \; n_{rd}(rd,buffer)$ \\
\indent $\quad \wedge \; write(n_{wr},data) \; \triangleright \; n_{wr}(wr,data)$ \\
$in \; \Pi_{n_i\in \overrightarrow{n}}(def \; n_{i}(c,p) \; | entry_i$$<$$sr,sw,se$$>$$ \; \triangleright $\\
\indent  \indent $(if \; [c = dl] \; then \; 0 \; else$\\
\indent  \indent $\: if \; [c = mv] \; then \; E(p,sr,sw,se) \; else$\\
\indent  \indent $\: if \; [c = ex] \; then \; se(p) \; | \; entry_i$$<$$sr,sw,se$$>$$ \; else$\\
\indent  \indent $\: if \; [c = rd] \; then \; p(sr()) \; | \; entry_i$$<$$sr,sw,se$$>$$ \; else$\\
\indent  \indent $\: if \; [c = wr] \; then \; sw(p) \; | \; entry_i$$<$$sr,sw,se$$>$$)$\\
\indent $in \; entry_i$$<$$sr_{i},sw_{i},se_{i}$ $\!\!>)$ \\[0,3em]
\end{small}

\noindent\textbf{Modeling the hierarchy of execution}\\
\indent The hierarchy of execution may vary from an operating system to an other, this introduces portability issues explaining that companion viruses gaining preemptive by modifying the hierarchy of execution are not very common \cite[Chpt.8]{FI05}. The most common case are companion viruses modifying the path variable in a Unix environment. An other example, a little bit outdated, concerns the DOS architecture where executable files with the \begin{small}\texttt{.com}\end{small} extension are preemptive on those with the \begin{small}\texttt{.exe}\end{small} extension. In fact, the hierarchy of execution relies on a shorter designation of programs (path or extension missing). These short designations are completed according to the hierarchy of execution. Let us first define a concatenation operator over names denoted $n_1\cdot n_2$ and a projection operator $\pi_n$ to recover the $n^{th}$ concatenated element. A process of completion must then be defined which is parametric over a list of complements (file path or extension), ordered by increasing preemptiveness:\\[0,3em]
\begin{small}
$H_{EX} \stackrel{\scriptscriptstyle \rm def}{=}$ \\
$complete(sn) \; | \; complist\!\!<\!\!c_0,...,c_n\!\!> \; \triangleright$\\
\indent  $let \; ln_0, ...,ln_n = sn\cdot c_0,...,sn \cdot c_n \; in$\\
\indent $(if \; [ln_0 \!\!\in\!\! dv] \; then \; return \; ln_0 \; | \; complist\!\!<\!\!c_0,...,c_n\!\!>$ \\
\indent  $else \; ... \; else$\\
\indent $if \; [ln_n \!\!\in\!\! dv] \; then \; return \; ln_n \; | \; complist\!\!<\!\!c_0,...,c_n\!\!>)$\\
$\wedge \; (preempt(c) \; | \; complist\!\!<\!\!c_0,...,c_n\!\!> \; \triangleright$\\
\indent $complist\!\!<\!\!c,c_0,...,c_{n-1}\!\!>$\\[0,3em]
\end{small}
\indent The execution command from the file system must be modified adequately to try name completion when the name of the program launched in execution is unknown from the system. In other words when the program name is not in the set of defined names.\\[0,3em]
\begin{small}
$M_{FS} \stackrel{\scriptscriptstyle \rm def}{=} def \; E_{FS} \; in$ \\
$def \; new(n_{new}) \; \triangleright \; E(n_{new},R_{exec}(null))$\\
\indent \indent $...$\\
\indent $\quad \wedge \; execute(n_{exe},arg) \; \triangleright$\\
\indent \indent \indent $if \; [n_{exec} \!\in\! dv] \; then \; n_{exe}(ex,arg)$\\
\indent \indent \indent $else \; execute(complete(n_{exec}),arg)$ \\
\indent \indent $...$\\
\end{small}

\noindent\textbf{Refining replication for companion viruses}\\
\indent From the Definition \ref{def:virus-classes}, the two classes of companion viruses can be obtained by refining the replication function $r$. Using this definition of a file system, a first companion virus $V$ diverting the file naming mechanism can be defined as follows:\\[0,3em] 
\begin{small}
$def \; r(v,n_{targ}) \; \triangleright $\\ $move(n_{targ},n_{copy});new(n_{targ});write(n_{targ},v) \; in \; ...$\\[0,3em]
\end{small}
The second class of companion viruses relies on the file system refining to support the execution hierarchy. Let us consider the target of the replication as a concatenated name $ln_{targ} = sn_{targ}\cdot ext$. The preemptive companion virus can be defined as follows:\\[0,3em] 
\begin{small} 
$def \; r(v,ln_{targ},ext) \; \triangleright $\\ 
$preempt(ext_{new});new(\pi_1(ln_{targ})\cdot ext_{new});$\\
$write(\pi_1(ln_{targ})\cdot ext_{new},v) \; in \; ...$\\[0,3em]
 \end{small}

\noindent\textbf{Model validation}\\
\indent In order to validate the model, it is necessary to assess its relevance with regards to existing companion viruses. A parallel has thus been drawn between the different processes and definitions, and their real implementation. A recent example of MacOS X virus circumventing the file naming mechanism has first been taken. The results are given in the Table \ref{tab:macosx}. The same work has been done for a second companion virus for Unix, diverting the execution hierarchy. The results are given in the Table \ref{tab:compunix}. 

\begin{table*}[!ht]
\centering
\begin{footnotesize}
\begin{tabular}{|l|l|}
\hline
\multicolumn{2}{|l|}{Companion Virus for Mac-0 Executables (\cite{FIL07l},2007)}\\
\multicolumn{2}{|l|}{Platform: Mac OS X}\\
\multicolumn{2}{|l|}{Type: Companion virus based on the directory structure of Mac-0 executables}\\
\hline
\rowcolor[gray]{0} \textcolor{white}{Processes} & \textcolor{white}{Implementation}\\
\hline
$M_{FS}$ & MacOS X file system with the Mac-0 executable structure in repositories: hierarchical tree and meta-information files.\\
\hline
$E_{FS}$ & $Info.plist$ containing information on the executable structure and the location of its elements.\\
\hline
\rowcolor[gray]{0} \textcolor{white}{Channels} & \textcolor{white}{Implementation}\\
\hline
$n_{targ}$ & The $CFBundledExecutable$ field from $Info.plist$ which denotes the real executable, the target of the infection.\\
\hline
$move$	& The $cp$ command from the console.\\
\hline
$create, write$ & The two commands are not detached and realized by a single call to the command $cp$.\\
\hline
\end{tabular}
\end{footnotesize}
\caption{Parallel with a Companion Virus for MacOS X based on file naming.}
\label{tab:macosx}
\end{table*}

\begin{table*}[!ht]
\centering
\begin{footnotesize}
\begin{tabular}{|l|l|}
\hline
\multicolumn{2}{|l|}{vcomp\_ex\_v1 (\cite[Chpt.8]{FI05},2005)}\\
\multicolumn{2}{|l|}{Platform: Unix}\\
\multicolumn{2}{|l|}{Type: Companion virus modifying environment variables for preemptiveness}\\
\hline
\rowcolor[gray]{0} \textcolor{white}{Processes} & \textcolor{white}{Implementation}\\
\hline
$M_{FS}$ & Unix file system.\\
\hline
$E_{FS}$ & Inode entries for the existing files.\\
\hline
$H_{EX}$ & The $PATH$ environment variable.\\
\hline
\rowcolor[gray]{0} \textcolor{white}{Channels} & \textcolor{white}{Implementation}\\
\hline
$n_{targ}$ & An absolute file name composed of the short file name and its path.\\
\hline
$preempt$	& The command $export \; PATH = NEW\_PATH:PATH$.\\
\hline
$create, write$ & The standard file API $fopen$ and $fwrite$.\\
\hline
\end{tabular}
\end{footnotesize}
\caption{Parallel with a Companion Virus for Unix based on execution hierarchy.}
\label{tab:compunix}
\end{table*}

\subsection{Stealth techniques inside Rootkits}

	Up until now, the article was only focusing on modeling self-replication since it is one of the main characteristics of malware and in particular viruses. In fact, the Join-Calculus is sufficiently expressive to describe other malicious behaviors such as stealth. Even if stealth is not a malicious technique on its own, deployed in rookits, it becomes a powerful tool for attackers. Unfortunately, few formal works have been led on rootkit modeling \cite{ZZ04,FIL07a,DV08}. Rootkits thus constitute an interesting choice to assess the expressiveness of the model, by proving it can be applied to concrete cases.\\
\indent This section describes how rootkit behaviors can be defined in the parametric model by refinement of the payload process which had not been detailed yet. Let us consider a piece of malware loading a rootkit from its body. Based on recursive functions, the definition published by Zuo and Zhou of viruses resident relatively to a system call is the closest result to our approach \cite{ZZ04}. Unfortunately, recursive functions are not really adapted to model reactive, persistent (non-terminating) programs such as rootkits. The Join-Calculus should offer far more flexibility.\\
	
\noindent\textbf{Services provided by the rootkit}\\
\indent Basically, a rootkit provides through a command channel a certain number of services to the attacker. Let us first define $n$ processes $S_1,...,S_n$ corresponding to these services. A public channel $com$ is provided to the attacker (through the network, based on various protocols such as IRC or P2P for the most spread). This channel supports $n$ different types of requests represented by the vector $\overrightarrow{c}=c_1...c_n$. The names $c_i$ themselves correspond to internal command channels, which, in the case of rootkits, are often communication channels from the user space where the client part is running, towards the services running in the kernel space. A service of proxy relays the commands received on the public channels towards the internal channels. This client-server architecture can be defined as follows. In the first place, a public communication channel $com$ must be defined between the attacker $A$ and the rootkit $R_{kit}$:\\[0,3em] 
\begin{small}
$D_{com} \stackrel{\scriptscriptstyle \rm def}{=} def \; com() \triangleright$\\
\indent $(def \; send \;$$<$$\overrightarrow{m}$$>$$|receive() \triangleright return \; \overrightarrow{m} \; to \; receive \; in$\\
\indent $return \; send,receive \; to \; com)\; in$\\
\indent $let \; sd,rcv = com() \; in \; (A|R_{kit})$\\[0,3em]
\end{small} 
In first place, the rootkit publishes the list of supported commands through the public channel. Once transmitted, it launches the proxy service waiting for requests from the attacker:\\[0,3em]
\begin{small}
$P_{proxy} \stackrel{\scriptscriptstyle \rm def}{=} let \; c,arg = rcv() \; in \; c(arg)$\\
$R_{kit} \stackrel{\scriptscriptstyle \rm def}{=} def \; c_1() \triangleright S_1 | P_{proxy}$\\
\indent \indent $\quad \wedge \; c_2(arg) \triangleright S_2 | P_{proxy}$\\
\indent \indent $\quad \wedge \; ...$ \\
\indent \indent $\quad \wedge \; c_n(arg) \triangleright S_n | P_{proxy} \; in \; sd$$<$$\overrightarrow{c}$$>$$.P_{proxy}$\\[0,3em]
\end{small}
In parallel, the attacker receives the available commands for the different services on the public channel. The obtained list is stored as the vector $\overrightarrow{s}$. He can then activate any service by sending a request containing the corresponding command: \\[0,3em]
\begin{small}
$A \stackrel{\scriptscriptstyle \rm def}{=} let \; \overrightarrow{s} = rcv() \; in \; sd$$<$$s_1,arg_1$$>$$.sd$$<$$s_2,arg_2$$>$$...$\\
\end{small}

\noindent\textbf{Loading the rootkit}\\
\indent Before installation, the rootkit is often stored in the malware body as an internal component. It must thus be extruded and loaded either conventionally through the driver manager or through a diverted mean. In both cases a specific loading process is required. Let us consider the conventional loading process by defining a service of driver manager. This service basically receives the driver definition and launches its execution:\\[0,3em]
\begin{small}
$D_{mdriv} \stackrel{\scriptscriptstyle \rm def}{=} load(d) \triangleright d\!\!<\!>$ \\[0,3em]
\end{small}
In order to be accessible inside the malware, the rootkit must be abstracted to ease the loading:\\[0,3em]
\begin{small}
$M \stackrel{\scriptscriptstyle \rm def}{=} (...);def \; r\!\!<\!>\triangleright R_{kit} \; in \; load\!\!<\!\!r\!\!>|M'$\\[0,3em]
\end{small}
The following loading process is obtained:\\[0,3em]
\begin{small}
$def \; D_{mdriv} \; in \; M \longrightarrow^* def \; G_{dr} \; in \; M'|R_{kit}$\\
\end{small}

\noindent\textbf{System call hooking}\\
\indent At last, it is necessary to model the hooking mechanism just like resident viruses in \cite{ZZ04}. Before, a new entity of the system must be defined: the system call table which is considered as a resource. This entity only publish the list of available system calls on-demand. This list is modeled by a vector of channel $\overrightarrow{sc}$ which can only be modified by the kernel through a privileged writing access. This privileged access is modified by the $hook$ channel which from the malware perspective is considered as private: only the $publish$ channel is returned at table creation:\\[0,3em]
\begin{small}
$D_{tsc} \stackrel{\scriptscriptstyle \rm def}{=} T_{sc}(\overrightarrow{t_{init}}) \; \triangleright $\\
\indent $def \; (publish() \; | \; table$$<$$\overrightarrow{t}$$>$$) \; \triangleright$\\
\indent \indent $(return \overrightarrow{t} \; to \; publish \; | \; table$$<$$\overrightarrow{t}$$>$$)$ \\
\indent $\!\wedge \, (hook(\overrightarrow{t_{new}}) \; | \; table$$<$$\overrightarrow{t}$$>$$) \; \triangleright(table$$<$$\overrightarrow{t_{new}}$$>$$)$ \\
\indent $in \; table$$<$$\overrightarrow{t_{init}}$$>$$ \; | \; return \; publish \; to \; T_{sc}$\\[0.3em]
\end{small}
To access to this privileged channel, the rootkit uses in a diverted way the system services and in particular the services of memory allocation. Allocation services can be used to modify the page protection of a memory space (\begin{small}\texttt{IoAllocateMdl}\end{small} under Windows \cite[pp.82-87]{HB06} and \begin{small}\texttt{Kmalloc}\end{small} under Linux \cite{SD01}). Generally speaking, allocation services take as input a base address $b$ and a size $s$ and return the result of the allocation. The $hook$ channel is only leaked if the base address is equal to the address of the system call table $scbase$. In any other case a simple acces is returned:\\[0,3em]
\begin{small}
$D_{alloc} \stackrel{\scriptscriptstyle \rm def}{=} alloc(b,s) \; \triangleright$\\
\indent $if \; [b\!=\!scbase] \; then \; return \; hook \; else \; return \; access$ \\[0,3em]
\end{small}
The interest of hooking for the rootkit is to define a set of false system calls $R_{fsc1}, ... , R_{fscm}$, in order to hide files or processes, for example by filtering the original system calls. These malicious system calls are registered in a new table which is a vector of $m$ entries $\overrightarrow{fsc}=fsc_1...fsc_m$ containing their referring names:\\[0,3em]
\begin{small}
$D_{fsc} \stackrel{\scriptscriptstyle \rm def}{=} def \; fsc_1(\overrightarrow{arg}) \; \triangleright \; R_{fsc1}$ \\
\indent $\wedge  \; ...$ \\
\indent $\wedge  \; fsc_m(\overrightarrow{arg}) \; \triangleright \; R_{fscm}$ \\
$R_{kit} \stackrel{\footnotesize \rm def}{=} def \; D_{fsc} \; in$ \\ 
\indent $let \; scspace = alloc(scbase,scsize) \; in \; scspace(\overrightarrow{fsc})$\\[0,3em]
\end{small}
The system evolves along the following derivation where the leak of the privileged writing channel is observed from the allocation mechanism:\\[0,3em]
\begin{small}
$def \; D_{tsc}\wedge D_{alloc} \; in \; let \; pub = T_{sc}(\overrightarrow{sc}) \; in \; R_{kit} \longrightarrow *$\\
\indent \indent \indent \indent \indent $\;\;\; def \; D_{tsc}\wedge D_{alloc} \wedge D_{fsc} \; in \; table$$<$$\overrightarrow{fsc}$$>$$ $  \\
\end{small}

\noindent\textbf{Model validation}\\
\indent In order to validate the model, it is necessary to assess its relevance with regards to existing rootkits. A parallel has thus been drawn between the different processes and definitions, and their real implementation in different malware. The results are given in the Tables \ref{tab:suckit}, \ref{tab:agony} and \ref{tab:agobot}. 
	
\begin{table*}[!ht]
\centering
\begin{footnotesize}
\begin{tabular}{|l|l|} 
\hline
\multicolumn{2}{|l|}{SuckIt (\cite{SD01,JAC07},2001)}\\
\multicolumn{2}{|l|}{Platform: Linux}\\
\multicolumn{2}{|l|}{Type: kernel space, system call hooking}\\
\hline
\rowcolor[gray]{0} \textcolor{white}{Processes} & \textcolor{white}{Implementation}\\
\hline
$M$ & $sk$, executable responsible for the rootkit installation from user space.\\
\hline
$R_{kit}$ & $core$, kernel module embedded in $sk$ to be loaded; it contains the provided services $S_n$. \\
\hline
$P_{proxy}$ & $backdoor$, autonomous thread waiting for network requests.\\
\hline
$D_{mdriv}$ & internal module of $sk$ responsible for allocating kernel memory, for writing the $core$ module,\\
 & and for resolving the addresses normally addressed by the $insmod$ command.\\
\hline
$D_{tsc}$ & Linux system call table.\\
\hline
$D_{alloc}$ & memory device $/dev/kmem$.\\
\hline
$R_{sc}$ &	hooked versions of the system calls $fork$, $open$, $read$, $kill$, ...\\
\hline
\rowcolor[gray]{0} \textcolor{white}{Channels} & \textcolor{white}{Implementation}\\
\hline
$com (sd, rcv)$ & established socket between the attacker and the $backdoor$ thread.\\
\hline
$\overrightarrow{c}$ & hooked version of the $olduname$ system call (kept for compability) allowing communication \\
 & between the $backdoor$ thread and the kernel module $core$ to transmit the different commands.\\
\hline
$load$ & calls to internal functions of $D_{mdriv}$.\\
\hline
$alloc$ & $kmalloc$.\\
\hline
$hook$ & write function called with the address returned by $kmalloc$.\\
\hline
$publish$ & $sysenter$ instruction allowing the switch between  user and kernel space according to the system call table.\\
\hline
$\overrightarrow{fsc}$ & calls to hooked functions through the replaced system call table.\\
\hline
\end{tabular}
\end{footnotesize}
\caption{Parallel with a Linux Kernel Rootkit: SuckIt.}
\label{tab:suckit}
\end{table*}

\begin{table*}[!ht]
\centering	
\begin{footnotesize}
\begin{tabular}{|l|l|}
\hline
\multicolumn{2}{|l|}{Agony (Sources available on the net by Intox7, 2006)}\\
\multicolumn{2}{|l|}{Platform: Windows}\\
\multicolumn{2}{|l|}{Type: kernel space, system call hooking}\\
\hline
\rowcolor[gray]{0} \textcolor{white}{Processes} & \textcolor{white}{Implementation}\\
\hline
$M$ & $agony.exe$, executable responsible for the rootkit installation from user space and for transmitting the commands.\\
\hline
$R_{kit}$ & $agony.sys$, kernel module embedded as a resource in $agony.exe$. Once loaded, it contains the different services $S_n$.\\
\hline
$P_{proxy}$ & $agony.exe$ transmits the keyboard input to the driver.\\
\hline
$D_{mdriv}$ & Windows Driver Manager called SCM ($Service Control Manager)$.\\
\hline
$D_{tsc}$ & SSDT Table ($System Service Descriptor Table$) containg the adresses of the Windows system calls.\\ 
\hline
$D_{alloc}$ & Memory allocation services.\\
\hline
$R_{sc}$ &	hooked versions of the system calls defined in the kernel module:\\
&  $ZwQuerySystemInformationHook$, $ZwQueryDirectoryFileHook$... \\
\hline
\rowcolor[gray]{0} \textcolor{white}{Channels} & \textcolor{white}{Implementation}\\
\hline
$com$ & Keyboard interface with the console application $Agony.exe$.\\
\hline
$\overrightarrow{c}$ & $DeviceIOControl$, a Windows system call used to communicate with drivers.\\
\hline
$load$ & Call to $CreateService$ followed by $StartService$.\\
\hline
$alloc$ & $MmCreateMdl$ now replaced by $IoAllocateMdl$.\\
\hline
$hook$ & Writing operation to the space newly allocated.\\
\hline
$publish$ & $sysenter$ instruction allowing the switch between  user and kernel space according to the system call table.\\
\hline
$\overrightarrow{fsc}$ & Adresses in memory of the new system calls defined in the kernel module.\\
\hline
\end{tabular}
\end{footnotesize}
\caption{Parallel with a Windows Kernel Rootkit: Agony.}
\label{tab:agony}
\end{table*}

\begin{table*}[!ht]
\centering	
\begin{footnotesize}
\begin{tabular}{|l|l|}
\hline
\multicolumn{2}{|l|}{AgoBot (\cite{AGO04}, first version in 2002)}\\
\multicolumn{2}{|l|}{Platform: Windows}\\
\multicolumn{2}{|l|}{Type: user space, hooking not supported}\\
\hline
\rowcolor[gray]{0} \textcolor{white}{Processues} & \textcolor{white}{Implementation}\\
\hline
$M$ & $Agobot$, originally a P2P worm, supporting in prior versions propagation through vulnerabilities.\\
\hline
$R_{kit}$ & $CBot$, C++ object defining the different services $S_n$ as well as their handlers.\\
\hline
$P_{proxy}$ & $CIrc$, objet C++ reponsable de la communication par IRC avec l'attaquant\\
\hline
$G_{dr}$ & $CInstaller$, C++ object responsible for copying the code and registering in the system (registry key).\\
\hline
\rowcolor[gray]{0} \textcolor{white}{Channels} & \textcolor{white}{Implementation}\\
\hline
$com$ & IRC communication established through the network.\\
\hline
$\overrightarrow{c}$	& call to the method $HandleCommand$ from the object $CBot$\\
\hline
$load$ & calls to the methods $CopyToSysDir$ and $RegSartAdd$ from the object $CInstaller$\\
\hline
\end{tabular}
\end{footnotesize}
\caption{Parallel with a Windows User Rootkit: Agobot.}
\label{tab:agobot}
\end{table*}

\section{System resilience / replication detection}

	Modeling facilities are not the only interest of the process algebra. Since the first formal works from Cohen, it is well established that virus detection is an undecidable problem. However, thanks to this formalism, we will now try to identify some fragments of the Join-Calculus for which the detection problem remains decidable up to a complexity factor. Let us consider an algorithm taking as input a system context $C_{sys}[\,.\,]_{S\cup R}$ and a process $P$ abstracted by the definition $p$. This algorithm returns true if $P$ is able to self-replicate inside the context.\\
\indent Such an algorithm can be used either for checking the process replication capability or assessing the context resilience to a viral class. An exhaustive procedure is described in the Algorithm \ref{alg-detect}. The purpose of this algorithm respectively changes whether the context or the tested process varies:
\begin{description}
\item[\textbf{Detection:}] $\quad\;$ Malware detection can be addressed by identifying replication attempts of various processes in a fixed system context.  
\item[\textbf{Resilience:}] $\quad\;\;$ Just like in any other domain of computer security, system resilience is addressed y confronting systems to different a known attack class. This problem can be addressed by identifying replication attempts of a given viral class in various system contexts. The viral class is defined through a fixed process in input, which is known to be a malware.
\end{description}

\begin{algorithm}
\caption{Replication detection. \label{alg-detect}}
\begin{small}
\begin{algorithmic}[1]
\REQUIRE $P$ which is abstracted by $p$\\
\REQUIRE $C_{sys}[\,.\,]_{S\cup R}$ where $S$ is the set of services and $R$ the resources\\
\STATE $E_{done} \leftarrow \oslash$, $E_{next} \leftarrow \oslash$, $C \leftarrow C_{sys}[P]_{S\cup R}$\\
\REPEAT
\STATE $E_{succ} \leftarrow \{C'|C\stackrel{\scriptscriptstyle \tau}{\longrightarrow}C'\}$\\
\IF{$\exists C'$ reached by a join pattern $x\!<\!p\!>$ with $x\in R$ or $x\not \in (dv(P) \cup S \cup R)$}
\STATE \textbf{return} \textit{system is vulnerable to the replication of $P$}\\
\ENDIF
\STATE $E_{done} \leftarrow E_{done} \cup \{C\}$\\
\STATE $E_{succ} \leftarrow E_{succ} \!\! \setminus \{C_d\!\! \in \! E_{succ}|\exists C_t \!\! \in \! E_{done}.C_d\equiv C_t \}$\\
\STATE $E_{next} \leftarrow E_{next} \cup E_{succ}$\\
\IF{infinite reaction on a join without apparition of new potential transitions}
\STATE \textbf{break}\\
\ENDIF
\STATE Choose a new $C\in E_{next}$\\
\UNTIL{$E_{next} \leftarrow \oslash$}
\STATE \textbf{return} \textit{system is not vulnerable to the replication of $P$} \\
\end{algorithmic}
\end{small}
\end{algorithm}

\begin{prop}\label{prop:undecidable}
Detection of self-replication in the Join-Calculus is undecidable.\\
\end{prop}

\begin{prop}\label{prop:decidable}
Detection of self-replication is decidable if the system context and the process are defined in the fragment of the Join-Calculus without name generation.\\
\end{prop}

Algorithm \ref{alg-detect} uses a brute-force approach for state exploration. As a matter of fact, it was not designed for operational deployment but to study the decidability of the detection problem. Without surprise, detection remains undecidable according to Proposition \ref{prop:undecidable}. However, according to Proposition \ref{prop:decidable}, the problem can become decidable by restricting name generation. This restriction is not without impact on the system context. Forbidding name generation induces a fixed number of resources without possibility to dynamically create new ones. But most importantly, without name generation, synchronous communication is no longer possible, in particular for services which can not generate fresh names to return values. Unique and fixed return channels must be specified instead.\\

\begin{IEEEproof}
\indent In the algorithm, the set of states $E_{succ}$ reached after a single reduction is finite because only internal transitions $\tau$ are considered. Internal transitions in join-calculus are finite state branching \cite{L96}. The decidability thus depends on the bounded number of iterations (finite number of states potentially reached and infinite loop detection). To prove the decidability, we will reduce the detection problem to the coverability problem in petri nets.\\
\indent Let us consider the fragment of the join-calculus without name generation i.e. no nested definitions of the form $def \; J\triangleright (def \; J'\triangleright P' \; in \; P) \; in \; Q$. This fragment can be encoded in the asynchronous $\pi$-calculus without external choice. Let us consider a similar encoding to \cite{CG96} except that the replication operator has been replaced by recursive equations in order to be consistent with the remaining of the proof: \\[0.5em]
\begin{small}
\noindent \begin{tabular}{r@{ }c@{ }l}
$[[Q|R]]_j$ & $=$ & $[[Q]]_j \; | \; [[R]]_j \qquad\quad$\\
$[[x\!\!<\!\!v\!\!>]]_j$ & $=$ & $\bar x v$\\
$[[def \; x\!\!<\!\!u\!\!>\!\!\,|\,y\!\!\,<\!\!v\!\!> \triangleright \, Q \; in \; R]]_j$ & $=$ & \\
\multicolumn{3}{r}{$\left \{
          \begin{array}{l} 
          A = x(u).y(v).([[Q]]_j \; | \;A)\\
         A \; |  \; [[R]]_j\\
     \end{array} \right \} $}\\
\end{tabular}\\
\end{small}

\normalsize
\indent Name generation being excluded and the process being considered in a close context, the scope restriction $\nu$ is absent from the encoding. We will now reuse the approach in \cite{AM01} to reduce the problem. Using the provided encoding, the process inside its system context can be encoded in the asynchronous $\pi$-calculus, resulting in a system of parametric equations satisfying the normalized form from \cite{AM01}.\\
\indent This system is then encoded into equations from the Calculus of Communicating Systems (CCS). CCS is parameterless, however, without name generation, channel $\sigma$  and possible transmitted value $a$ can be combined in a single channel $<\!\!\sigma , a\!\!>$. Notice that this encoding reintroduces the external choice $+$ to handle the combined channels. Just like in \cite{AM01}, the obtained equation system thus contains a set of parallel processes guarded by these channels. The only differences lie in the multiple join patterns in join-calculus which results in multiple channels guarding these processes:
\begin{center}\begin{small} $A_i = \Sigma <\!\!\sigma , a\!\!>.<\!\!\sigma' , a'\!\!>.(\Pi \; \overline{<\!\!\sigma , a\!\!>} \;|\; \Pi \; A_j)$ \end{small}\end{center} 

\indent In this equation system, the replication detection is reduced to the problem of knowing if one of the guarded process $A_i$ is activated over a channel $<\!\!\sigma , p\!\!>$ with $\sigma \in R$ and $p$ is the abstraction of $P$. This is typically a control reachability problem in CCS. It has been proven in \cite{AM01} that control reachability can finally be reduced to the coverability problem in petri nets. Although it is time and space consuming, there exist decidable algorithms computing coverability \cite{KU05} and thus able to detect any token in the $\sigma p$ place, referring to the emission of the process definition on the $\sigma$ the channel.
\end{IEEEproof}

\section{Policies to prevent malware propagation}
	The previous section deals with the problem of malware detection through their self-replication characteristic. It has been proven that detection was decidable only under certain assumptions. The problems concerning decidability and the fact that detection is reactive and not proactive encourage the research of alternative solutions to fight against malware. It is thus important to consider other proactive approaches such as the prevention of malware propagation. This section first describes the malware propagation as an illegal information flow and then envisage different solutions for malware containment.
	
\subsection{Non-infection property and isolation}
\label{subsec:non-infect}
	A different approach to fight back the threat brought by malware is to reason in terms of information flow as initiated by F. Cohen in \cite{CO87}. Active research works are currently led in order to control illicit data flows between processes of different security levels \cite{GM82,RS02,HR02c}. One of the main result is the formalization of the non-interference property which specifies that the behavior of a low-level process must not be influenced by an upper-level process. This non-interference property is used for addressing confidentiality issues.\\
\indent Similarly, the replication process of malware can be compared to an illicit information flow of the viral code towards the system. Let us state the hypothesis that, contrary to malware, legitimate programs should not interfere with other processes implicitly through the system. This is a typically an integrity issue, and the non-interference property must be adapted accordingly. We have thus defined in Theorem \ref{theo:non-infection} a new property called non-infection in reference to the original property of non-interference \cite{GM82}. \\

\begin{theo} (NON-INFECTION). \label{theo:non-infection} Let us consider a process $P$ placed into a system context considered stable (i.e. potential reactions to intrusions only). The property of non-infection is satisfied by $P$ if the system evolves along the reaction $C_{sys}[P]\longrightarrow^* C'_{sys}[P']$, and for any non-infecting test process $T$ the equivalence $C_{sys}[T]\approx C'_{sys}[T]$ is true. The strength of the property is determined by the equivalence considered.\\
\end{theo}

\indent The non-infection property guarantees the integrity of the system context. With regards to this property, the consequent question is to know what are the mandatory constraints for a system context to satisfy non-infection. The Proposition \ref{prop:isolation} states that there exist systems preventing replication through resource isolation. This proposition in fact corresponds to a generalization of the network partitioning principle advocated by F. Cohen to fight virus propagation \cite{CO87}. \\

\begin{prop} \label{prop:isolation} In a system context made up of services and resources, the non-infection property can only be guaranteed by a tight isolation of the resources. \\
\end{prop}

\begin{IEEEproof}
Let us consider a system context made up of services and resources (see Section \ref{sec:environment}) of the form: $C_{sys} = def \; D_S \wedge D_R \; in \; R \; | \; [\,.\,]$\\
By hypothesis the context is stable and will only react to intrusions from the process $P$ placed inside. To prove that isolation is required, we show that any writing access to a resource, either direct or indirect, must be forbidden. Let us begin by enumerating the possible intrusion cases from the process $P$:\\[0,5em]
\noindent \textbf{I. Intrusion towards a resource:}\\
\noindent $J \in D_R$ with $J=x_1(\overrightarrow{y_1})|...|x_{n}(\overrightarrow{y_n})\triangleright R'$\\

\begin{centering} \begin{small} $def \, D_S \wedge D_{R} \! \setminus \! \{J\} \wedge J \; in \;  R_0|x_1(\overrightarrow{z_1}).R_1|...|x_{m}(\overrightarrow{z_m}).R_m|[\,.\,]$\\
$\xrightarrow{\qquad \scriptscriptstyle x_{m+1}(\overrightarrow{z_{m+1}})|...|x_{n}(\overrightarrow{z_n})\qquad}$\\
$def \; D_S \wedge D_R \; in \; R_0|R_1|...|R_m|R'[\overrightarrow{y}/\overrightarrow{z}]|[\,.\,]$.\\[0,5em]
\end{small} \end{centering}
\noindent This can be simplified since in our model the $x_i$ are only used to store the resource content meaning that $R_i=0$ for $1 \! \leq \! i \leq \! m$. From there, there are three sub-cases for this transition.\\[0,2em]
\textit{1) Reading from the resource:}\\
\begin{small}$R'\equiv x_1(\overrightarrow{y_1})|...|x_{m}(\overrightarrow{y_m}) \;| return \; \overrightarrow{y_1},...,\overrightarrow{y_m} \; to \; x_{m+1}$\end{small}. Once the return consumed, the system recover its initial state before the intrusion: the non-infection property is satisfied. \\[0,2em]
\textit{2) Writing to the resource:}\\
\begin{small}$R'\equiv x_1(\overrightarrow{y_{m+1}})|...|x_{m}(\overrightarrow{y_n} )| return \; to \; x_{m+1}$\end{small}. Once the return consumed, the original values $y_i$ with $1 \! \leq \! i \! \leq \! m$ are sustituted by values $y_j$ with $m+1 \! \leq \! j \! \leq \! n$. It is not thus guaranteed that the system will recover its original state before the intrusion: the non-infection property may not be satisfied.\\[0,2em]
\textit{3) Executing the resource:}\\
This sub-case is equivalent to intrusion towards a service (see II.).\\

\noindent \textbf{II. Intrusion towards a service:}\\
$J \in D_S$ with $J=x_1(\overrightarrow{y_1})|...|x_{n}(\overrightarrow{y_n})\triangleright S$\\

\begin{centering} \begin{small} $def \; D_S\setminus \{J\} \wedge J \wedge D_R  \; in \; R \; | \; [\,.\,]$\\
$\xrightarrow{\qquad \scriptscriptstyle x_{1}(\overrightarrow{z_{1}})|...|x_{n}(\overrightarrow{z_n}) \qquad}$\\
$def \; D_S \wedge D_R \; in \; S[\overrightarrow{y}/\overrightarrow{z}] \; | \; R' \; | \; [\,.\,]$\\[0,5em]
\end{small}
\end{centering}
\noindent $S$ is of the form $return \; f(\overrightarrow{z_1},...,\overrightarrow{z_n}) \; to \; x_{1}$ which reduces to the null process when the return is consumed. The system modification thus depends on the nature of the fonction $f$. Once again, there are three sub-cases.\\[0,2em]
\textit{1) Definition of $f$ accessing no resource or only through a reading channel:} This case is identical to the case I.1) and the non-infection property is satisfied. \\[0,2em]
\textit{2) Definition of $f$ using a writing or creation channel for resources:} This case is identical to the case I.2) and the non-infection property may not be satisfied. \\[0,2em]
\textit{3) Definition of $f$ accessing resources in execution:} In this case, the solution depends on the content of the resource. The same test is applied recursively to this content until reaching the cases II.1) or II.2).
\end{IEEEproof}

\subsection{Policies to restrict infection scope}
	The non-infection property is impossible to guarantee in practice. The complete isolation of resources can not obviously be considered in systems without loosing most of their use \cite{CO87}. In fact, the hypothesis stated in \ref{subsec:non-infect} about legitimate programs is not always true in real cases. But if non-infection is impossible to deploy, approximate solutions can still contain the malware propagation by restricting spatially and temporally the resource accesses. Such a restriction does not completely prevent malware propagation but the scope of the propagation is at least be confined. \\
\indent Such a restriction can be deployed by an access authority, blocking any unauthorized access to the resources and services of a system. A solution based on access tokens can be considered, either for spatial restriction (only program and resources sharing the same token can access each other) or for time restriction in terms of counting executions (a given token can be used a fixed number of times). As defined in \cite{YPG00}, an access authority is generically made up of two components: a Policy Decision Point (PDP) which can be seen as the token distribution mechanism and a Policy Enforcement Point (PEP) which checks the token validity and thus must not be bypassed (Definition \ref{defi:contol-access}). The obligation to pass through a verification authority is similar to the transitive non-interference where high-level information can only transit to low-level channels through an intermediate \cite{RS02}. This is reserved for future works.\\

\begin{defi}\label{defi:contol-access} An access authority is constituted of:\\
-A distribution process to deliver tokens denoted $D_T$.\\
-A control mechanism providing interfaces $\overrightarrow{chk}$ to submit tokens for checking.\\
The interfaces and the control mechanism are directly embedded in the system. The control is securely enforced (i.e. can not be bypassed) if the system without the distribution process satisfies the non-infection property.\\
\end{defi}

\begin{exa}
Let $T$ be a security token, non-forgeable i.e. if unknown, the token can not be rebuilt. $T$ must thus not be exported by the system context: $C_{sys}[.]_{S\cup R\cup {\scriptscriptstyle \overrightarrow{chk}}}$ with $T \not \in S$ and $T \not \in R$. Control can then be enforced at the resources and services level using the interface $chk$ which compare the token in entry with the security token $T$:\\[0,5em]
\begin{small}
$\bullet  \; def \; S_{sv}(t,\overrightarrow{arg}) \; \triangleright$\\
\indent $if \; chk(t,T) \; then \; return \; f_{sv}(\overrightarrow{arg}) \; else \; 0 \; in \; ...$\\[0,5em]
\noindent $\bullet \; def \; R_{exec}(f_0) \; \triangleright $\\
$\; def \; (write(t,f_{new})|content$$<$$f$$>$$) \; \triangleright$\\
$\; if \; chk(t,T) \; then \; (return \; to \; write |content$$<$$f_{new}$$>$$)$\\
\indent \indent \indent \indent \indent $else \; content$$<$$f$$>$$ $\\
$\wedge \; (read(t)|content$$<$$f$$>$$) \; \triangleright$\\
$if \; chk(t,T) \; then \; (return \; f \; to \; read|content$$<$$f$$>$$)$\\
\indent \indent \indent \indent \indent $else \; content$$<$$f$$>$$ $\\
$\wedge \; (exec(t,\overrightarrow{arg})|content$$<$$f$$>$$) \; \triangleright$\\
$if \; chk(t,T) \; then \; (return \; f(\overrightarrow{arg}) \; to \; exec|content$$<$$f$$>$$)$\\
\indent \indent \indent \indent \indent $else \; content$$<$$f$$>$$ $ \\
$in \; content$$<$$f_0$$>$$|return \; read,write,exec \; to \; R_{exec} \; in \; ...$\\
\end{small}
\end{exa}
	
\begin{figure*}[!ht]
  \centering
  \includegraphics*[scale=0.32]{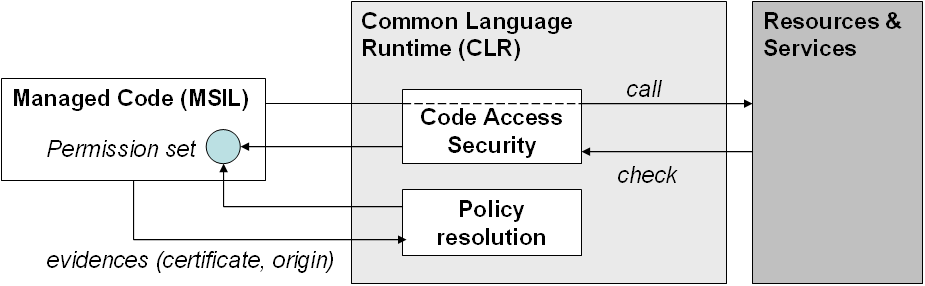} 
  \caption{.NET Security Model for access control.}
  \label{fig:dotnet}
\end{figure*}

\indent The example above is quite basic. It shows that if $T$ is not forgeable and no distribution mechanism is responsible for its extrusion, the process placed in the context will not be able to access any service and resource. Mechanisms of access control definitely help to contain malware propagation. In fact, complete access control mechanims are already deployed in two well known security models for Java \cite{java} and .Net \cite{dotnet}. In both models, the managed code is run in a isolated runtime environment (Java Virtual Machine or Common Language Runtime) with a controlled access to resources. A schematic view of the access control in the .Net framework is given in Figure \ref{fig:dotnet}. A parallel between the two models is given in the table below. These two access control models are already used to restrict malware propagation by restraining the number of services and resources available to untrusted codes. For example, the Same Origin Policy (SOP) forbid accesses to local resources, to any remote code running inside a web-browser.  The problem in actual system is that these controls are restricted to managed language and not to native code. These works on malware prevention prove that extending control to native code would help to fight malware propagation.\\

\noindent	
\begin{footnotesize}
\begin{tabular}{|@{ }l@{ }|@{ }l@{ }|@{ }l@{ }|}
\hline
\rowcolor[gray]{0} \textcolor{white}{Model} & \textcolor{white}{Java framework}  & \textcolor{white}{.NET framework}\\
\hline
Token distribution & Secure class & Policy resolution\\
(process $D$) & loader & of the Common\\
 &  & Language Runtime\\
 &  & (CLR)\\
\hline
Input for & Evidences & Evidences \\
distribution & (certificate, origin) & (certificate, origin)\\
\hline
Output & Permission domain & Permission set\\
(token $T$) &  & \\
\hline
Access control & Security Manager & Code Access\\
(interface $chk$) & calling the Access  & Security (CAS)\\
 & Controler using & enforced by\\
 & Checkpermission() & the CLR\\
\hline
\end{tabular}
\end{footnotesize}

\section{Conclusion and perspectives}

\indent This paper introduces the basis for a unified malware model based on process algebra and more particularly the Join-Calculus. Moving from the functional models currently used in abstract virology to process-based models do not result in a loss of expressiveness. The fundamental results are supported by the new model: characterization of the self-replication, undecidability of the detection and isolation as perfect prevention.

\indent In addition, the new model offers a greater expressiveness by the support of interactions, concurrency and non-termination which are commonly used in recent malware. In addition to computational aspects, these interactive notions ease the definition of complex behaviors such as stealth in rootkits. But modeling is not the only benefit; use of process algebra has provided new fundamental results in terms of detection and prevention. Even if the global problem of virus detection remains undecidable in this formalism, a fragment of the Join-Calculus where detection becomes decidable has been identified. With regards to prevention, the property of non-infection has been precisely defined as well as solutions to restrict malware propagation.

\indent In fact, just like no-interference, non-infection is a property which proves too strong for real cases. Approximate solutions based on security tokens have been evoked in the paper but future works can be led to reduce the strength of the property. Looking at existing works in process algebra, a promising perspective is to associate security levels to process through a typing mechanism. \\

\section{Works in progress}

\subsection{Security levels and typing}

\begin{theo} (RESTRICTED NON-INFECTION). \label{theo:rest-non-infection} Let us consider a process considered with potentail risk $\Gamma \Vdash^{risk}P$ placed into a system context considered stable (i.e. potential reactions to intrusions only) and legitimate $\Gamma \Vdash^{leg} C_{sys}[0]$. The property of non-infection is satisfied by $P$ if the system evolves along the reaction $C_{sys}[P]\longrightarrow^* C'_{sys}[P']$, and for any non-infecting test process $\Gamma \Vdash^{leg} T$ the equivalence $C_{sys}[T]\approx C'_{sys}[T]$ is true. The strength of the property is determined by the equivalence considered.\\
\end{theo}

Just like the original non-infection, restricted non-infection is only achieved if a complete isolation is made between legitimate and risky resources. This property is less strong and allows the modification of legitimate resources between them. Other typings may be defined as controls either for resources accesses (parallel with behavioral blocking) or information flows (parallel with tainting techniques) to prevent self-replication.

\subsection{Stealth and observation}

Let $O$ be a process monitoring one or several behaviors $\mu = \mu_1...\mu_n$ in a system $S$ (IDS or behavioral AV). When an attack is detected, the process switches to a state signaling the detection $O_I$:\\
$O|S\stackrel{\scriptscriptstyle \mu_i}{\longrightarrow}O_I|S'$.\\

\begin{defi} (STEALTH) Let us define stealth relatively to an observer (and no longer to a system call \cite{ZZ04}). The definition of an observer determines the observed behaviors which may bay coumpound of system calls. A malicious code $M$ is stealthy with respect to an observer $O$ if:\\
$O|S|M\not\longrightarrow ^* O_I|S'$.\\
\end{defi}

A malicious code can be stealthy for any legitimate observer. However, malware necessarily modify legitimate programs, otherwise the non-infection property would be satisfied and it would not be a malware. An observer can thus be found to detect a malware. In other words \textbf{absolute stealth for malicious code is impossible}. This result is promising for behavioral observation. A parallel can be drawn with E. Filiol's result saying that it is not possible to introduce a stealthy malicious code without modifying significantly the distribution for an estimator  \cite{FIL07a}.\\

\thanks{\textbf{Acknowledgement:} The author would like to thank particularly Guillaume Bonfante and Jean-Yves Marion from the LORIA for their precious help during the exploration of the different process algebras. Their working leads on the decidability of the detection as well as token-based solutions to propagation restriction have led to interesting results. The author would also like to thank Eric Filiol for his discerning parallels between the process model and existing works in virology, as well as Hervé Debar for its relevant parallels with concrete issues such as rootkit codes and access control architectures.\\}

\thanks{\textbf{Acknowledgement:} This work has been partially supported by the European Commissions through project FP7-ICT-216026-WOMBAT funded by the 7th framework program. The opinions expressed in this paper are those of the authors and do not necessarily reflect the views of the European Commission.}	

\bibliography{malware_process_algebra}

\begin{thebibliography}{10}
\providecommand{\url}[1]{#1}
\csname url@samestyle\endcsname
\providecommand{\newblock}{\relax}
\providecommand{\bibinfo}[2]{#2}
\providecommand{\BIBentrySTDinterwordspacing}{\spaceskip=0pt\relax}
\providecommand{\BIBentryALTinterwordstretchfactor}{4}
\providecommand{\BIBentryALTinterwordspacing}{\spaceskip=\fontdimen2\font plus
\BIBentryALTinterwordstretchfactor\fontdimen3\font minus
  \fontdimen4\font\relax}
\providecommand{\BIBforeignlanguage}[2]{{%
\expandafter\ifx\csname l@#1\endcsname\relax
\typeout{** WARNING: IEEEtran.bst: No hyphenation pattern has been}%
\typeout{** loaded for the language `#1'. Using the pattern for}%
\typeout{** the default language instead.}%
\else
\language=\csname l@#1\endcsname
\fi
#2}}
\providecommand{\BIBdecl}{\relax}
\BIBdecl

\bibitem{CCGKP08}
L.~Cardelli, E.~Caron, P.~Gardner, O.~Kahramanogullari, and A.~Phillips, ``A
  process model of actin polymerisation,'' in \emph{Proceedings of the From
  Biology to Concurrency and Back Conference (FBTC'08), Satellite Workshop of
  ICALP}, 2008.

\bibitem{KN06}
C.~Kuttler and J.~Niehren, ``Gene regulation in the $\pi $-calculus: Simulating
  cooperativity at the lambda switch,'' \emph{Transactions on Computational
  Systems Biology}, vol. Lecture Notes in Bioinformatics 4230, no. VII, pp.
  24--55, 2006.

\bibitem{JFD07}
G.~Jacob, E.~Filiol, and H.~Debar, ``Malwares as interactive machines: A new
  framework for behavior modelling,'' \emph{Journal in Computer Virology},
  vol.~4, no. 3, DIMVA and TCV Special Issue U. Flegel, G. Bonfante and J-Y.
  Marion Eds., pp. 235--250, 2008.

\bibitem{AD90}
L.~M. Adleman, ``An abstract theory of computer viruses,'' in \emph{CRYPTO '88:
  Proceedings on Advances in cryptology}, 1990, pp. 354--374.

\bibitem{FI05}
E.~Filiol, \emph{Computer viruses: from theory to applications}.\hskip 1em plus
  0.5em minus 0.4em\relax Springer, IRIS Collection, 2005.

\bibitem{BKM06}
G.~Bonfante, M.~Kaczmarek, and J.-Y. Marion, ``On abstract computer virology
  from a recursion-theoretic perspective,'' \emph{Journal in Computer
  Virology}, vol.~1, no. 3-4, pp. 45--54, 2006.

\bibitem{FI07b}
E.~Filiol, ``Formalisation and implementation aspects of k-ary (malicious)
  codes,'' \emph{Journal in Computer Virology}, vol.~3, no. 3, EICAR 2007 Best
  Academic Papers, V. Broucek and P. Turner Eds., 2007.

\bibitem{FIL07a}
------, ``Formal model proposal for (malware) program stealth,'' in
  \emph{Proceedings of the Virus Bulletin Conference (VB2007)}, 2007.

\bibitem{DV08}
A.~Derock and P.~Veron, ``Another formal proposal for stealth,'' in
  \emph{Proceedings of World Academy of Science, Engineering and Technology
  (WASET)}, 2008, pp. 158--164.

\bibitem{M99}
R.~Milner, \emph{Communicating and Mobile Systems: the $\pi $-calculus}.\hskip
  1em plus 0.5em minus 0.4em\relax Cambridge University Press, 1999.

\bibitem{FO98}
C.~Fournet, ``The join-calculus: a calculus for distributed mobile
  programming,'' Ph.D. dissertation, Ecole Polytechnique, Palaiseau, November
  1998.

\bibitem{FG00}
C.~Fournet and G.~Gonthier, ``The join calculus: a language for distributed
  mobile programming,'' in \emph{Draft 7/01, Applied Semantics Summer School
  (Caminha)}, 2000, pp. 1--66.

\bibitem{vN66}
J.~von Neumann, \emph{Theory of Self-Reproducing Automata}.\hskip 1em plus
  0.5em minus 0.4em\relax University of Illinois Press, 1966.

\bibitem{KRA80}
J.~Kraus, ``Selbstreproduktion bei programmen, {Ph.d. Dissertation at the
  University of Dortmund in 1980}, translated and edited by {D. Bilar} and {E.
  Filiol} under the title {O}n self-reproducing programs,'' \emph{Journal in
  Computer Virology}, vol.~5, no.~1, 2009.

\bibitem{Rog87}
H.~Rogers, \emph{Theory of Recursive Functions and Effective
  Computability}.\hskip 1em plus 0.5em minus 0.4em\relax The MIT Press, 1987.

\bibitem{WM07}
M.~Webster and G.~Malcolm, ``Reproducer classification using the theory of
  affordances,'' in \emph{Proceedings of the IEEE Symposium on Artificial Life
  (CI-ALife)}, 2007, pp. 115--122.

\bibitem{CO89}
F.~B. Cohen, ``Computational aspects of computer viruses,'' \emph{Computers \&
  Security}, vol.~8, no.~4, pp. 325--344, 1989.

\bibitem{FIL07l}
E.~Filiol, ``Mac os x n'est pas invulnérable aux virus : comment un virus se
  fait compagnon,'' \emph{Linux Magazine, Virus UNIX, GNU/Linux \& Mac OS X},
  pp. 20--31, September 2007.

\bibitem{ZZ04}
Z.~Zuo and M.~Zhou, ``Some further theoretical results about computer
  viruses,'' \emph{The Computer Journal}, vol.~47, no.~6, pp. 627--633, 2004.

\bibitem{HB06}
G.~Hoglund and J.~Butler, \emph{Rootkits, Subverting the Windows kernel}.\hskip
  1em plus 0.5em minus 0.4em\relax Addison-Wesley Professional, 2006.

\bibitem{SD01}
Sd and Devik, ``0x07 - linux on-the-fly kernel patching without lkm,''
  \emph{Phrack}, vol.~58, 2001.

\bibitem{JAC07}
G.~Jacob, ``Technologie rootkit sous linux/unix,'' \emph{Linux Magazine, Virus
  UNIX, GNU/Linux \& Mac OS X}, pp. 47--57, September 2007.

\bibitem{AGO04}
\BIBentryALTinterwordspacing
infectionvectors.com, ``Agobot and the "kit"-chen sink,'' 2004. [Online].
  Available: \url{www.infectionvectors.com/vectors/kitchensink.htm}
\BIBentrySTDinterwordspacing

\bibitem{L96}
C.~Laneve, ``May and must testing in the join-calculus: Ublcs-96-4,''
  University of Bologna, Tech. Rep., 1996.

\bibitem{CG96}
C.~Fournet and G.~Gonthier, ``The reflexive cham and the join-calculus,'' in
  \emph{In Proceedings of the 23rd ACM Symposium on Principles of Programming
  Languages}.\hskip 1em plus 0.5em minus 0.4em\relax ACM Press, 1996, pp.
  372--385.

\bibitem{AM01}
R.~M. Amadio and C.~Meyssonnier, ``On decidability of the control reachability
  problem in the asynchronous $\pi$-calculus,'' \emph{Nordic Journal of
  Computing}, vol.~9, no.~2, pp. 70--101, 2002.

\bibitem{KU05}
P.~Küngas, ``Petri net reachability checking is polynomial with optimal
  abstraction hierarchies,'' in \emph{In Proceedings of the 6th International
  Symposium on Abstraction, Reformulation and Approximation (SARA)}, 2005.

\bibitem{CO87}
F.~B. Cohen, ``Computer viruses: Theory and experiments,'' \emph{Computers \&
  Security}, vol.~6, no.~1, pp. 22--35, 1987.

\bibitem{GM82}
J.~A. Goguen and J.~Meseguer, ``Security policies and security models,'' in
  \emph{Proccedings of the Symposium on Security and Privacy (SSP'82)}.\hskip
  1em plus 0.5em minus 0.4em\relax IEEE Computer Society Press, 1982, pp.
  11--20.

\bibitem{RS02}
P.~Y.~A. Ryan and S.~A. Schneider, ``Process algebra and non-interference,''
  \emph{Journal of Computer Security}, vol.~9, no. 1-2, pp. 75--103, 2001.

\bibitem{HR02c}
M.~Hennessy and J.~Riely, ``Information flow vs. resource access in the
  asynchronous pi-calculus,'' \emph{ACM Transactions on Programming Languages
  and Systems}, vol.~25, no.~4, pp. 566--591, 2002.

\bibitem{YPG00}
R.~Yavatkar, D.~Pendarakis, and R.~Guerin, ``A framework for policy-based
  admission control,'' RFC-2753, Tech. Rep., 2000.

\bibitem{java}
L.~Gong, G.~Ellison, and M.~Dageforde, \emph{Inside Java$^{TM}$ 2 Platform
  Security: Architecture, API Design, and Implementation (2$^{nd}$
  Edition)}.\hskip 1em plus 0.5em minus 0.4em\relax Prentice Hall PTR, the Java
  Series, 2003.

\bibitem{dotnet}
A.~Freeman and A.~Jones, \emph{Programming .NET Security}.\hskip 1em plus 0.5em
  minus 0.4em\relax O'Reilly, 2003.

\end{thebibliography}
\bibliographystyle{ieeetran}

\end{document}